\documentclass[a4paper,12pt,authoryear]{article}
\usepackage{amssymb,amsmath,mathtools,amsthm} 
\usepackage{enumitem,parcolumns,multirow}
\usepackage[titletoc,toc,title]{appendix}
\usepackage{subfigure}
\usepackage{tikz,pgfplots,pgfplotstable}

\newcommand{\abs}[1]{\lvert #1 \rvert}

\theoremstyle{remark}

\theoremstyle{theorem}

\newlength\figureheight 
\newlength\figurewidth
\usepackage{authblk}
\usepackage[a4paper, total={6.5in, 9in}]{geometry}
\usepackage{lscape}

\pgfplotsset{compat=1.12}
\begin{document}
\author[1]{Elisa Letizia}
\author[1,2,3]{Fabrizio Lillo}
\affil[1]{Scuola Normale Superiore, piazza dei Cavalieri 7, 56126, Pisa, Italy - elisa.letizia@sns.it}
\affil[2]{Department of Mathematics, University of Bologna, Piazza di Porta San Donato 5, 40126, Bologna, Italy - fabrizio.lillo@unibo.it}
\affil[3]{Center for Analysis, Decisions, and Society, Human Technopole, Milano, Italy}
\setcounter{Maxaffil}{0}
\renewcommand\Affilfont{\itshape\small}

\title{Corporate payments networks\\ and credit risk rating}
\date{}

\maketitle
\begin{abstract}
Aggregate and systemic risk in complex systems are emergent phenomena depending on two properties: the idiosyncratic risks of the elements and the topology of the network of interactions among them. While a significant attention has been given to aggregate risk assessment and risk propagation once the above two properties are {\it given}, less is known about how the risk is distributed in the network and its relations with the topology. We study this problem by investigating a large proprietary dataset of payments among 2.4M Italian firms, whose credit risk rating is known. We document significant correlations between local topological properties of a node (firm) and its risk. Moreover we show the existence of an homophily of risk, i.e. the tendency of firms with similar risk profile to be statistically more connected among themselves. This effect is observed when considering both pairs of firms and communities or hierarchies identified in the network. We leverage this knowledge to show the predictability of the missing rating of a firm using only the network properties of the associated node. 

\vspace{5pt}
\textit{keyword}
  financial networks - corporate networks - credit risk - credit rating - machine learning

  
\end{abstract}

Assessing the aggregate risk emerging in complex systems is of paramount importance in disparate fields, such as economics, finance, epidemiology, infrastructure engineering, etc.. A large body of recent literature has explored, both theoretically and empirically, how risk propagates \cite{pozzi2013spread} and how to assess aggregate risk  when the risk of each individual entity is known \cite{nier2007network}, as well as the topology of the network of interaction among them. Both aspects have been shown to be important, however their mutual relation is relatively less explored. In theoretical studies, one typically assumes independence between idiosyncratic risk and topology, while in empirical studies the correlation is the one present in the investigated dataset.

But what is the relation (if any) between the idiosyncratic risk of a node and its local topological properties (e.g. degree, centrality, community, etc)? In this paper we answer this question by studying a specific system where the assessment of aggregate risk is particularly important, namely the network of interaction between firms. Assessing the risk of firms is one of the fundamental activities of the credit system. Banks spend a significant amount of resources to scrutinise the balance sheet of firms in order to obtain accurate estimations of their riskiness, the internal rating, and provide credit conditions reflecting both the capability of the firm to repay the loans and its probability of default. The riskiness of a firm depends on many idiosyncratic factors (e.g. balance sheet, structure of management, etc.) as well as the industrial sector or its geographical location \cite{treacy2000credit,crouhy2000comparative,Crouhy200147}. However, corporate firms do not live in isolation but interact with each other on a daily basis. The interactions can be of different kinds, including those due to the supply chain, payments, business partnerships, financial contracts, and mutual ownership. The structure of interactions is complex and multifaceted but its knowledge is critical both for macroeconomists and for the credit and banking industry to understand the dynamics of the economy, the business cycle, the structure of corporate control, and, of course, the risk of firms (in isolation or in aggregation).

Here we study the interplay between the risk of firms and the interlinkages connecting them. The network is built from a large proprietary dataset provided by a major European bank. The dataset contains the payments collected at daily granularity between more than two million Italian firms together with the information on internal risk rating for a large fraction of them. 
We want to understand whether and in which measure a firm's role in the network can be informative of its riskiness. This is important for two reasons. First, even if the risk of a firm is not known to all the counterparts, it may affect its ability to interact with other firms. For example, a poor rating (i.e. high riskiness) may prevent the access to credit and as a result it may cause a reduction or delay in payments toward suppliers. If the supplier has high risk, the missing or delayed payment can prevent its own payments, increasing the likelihood of a cascade of missing payments and a propagation of financial distress. The second reason is that, in certain cases, the knowledge of the riskiness of a firm or a group of firms is lacking or imprecise. In these cases, the existence of a correlation between network properties and risk can allow or improve the assessment of risk. Indeed, in the last part of the paper we will show how network properties of a node can be used to predict the risk of the corresponding firm. 

Previous works on networks of firms focussed mainly on ownership \cite{kogut2001small, souma2006change, vitali2011network, Romei2015,garcia2017uncovering}, or dealt with the modelling but not with the empirical part \cite{huremovic2016production}. Exception are the empirical studies on the the Japanese economic system \cite{takaaki2009, watanabe2012}, where links represent buyer-supplier relationship between, but without the information on the amount of money exchanged. In the seminal work \cite{acemoglu2012network}, even if the theoretical framework applies also to single firms, the empirical part focuses on the aggregate, sector network, due to lack of more granular data. The use of payments as a proxy of interactions between economic entities is not new and has been investigated mainly for banks \cite{soramaki2007topology, rordam2009topology,battiston2012debtrank, bargigli2015multiplex, fukuyama2016modelling} in the context of systemic risk studies, where, however, other choices to characterise interactions are possible \cite{elliot2014financial,cimini2015systemic,affinito2017interbank, d2017does}. Apparently much less is known about the payment network between firms, mostly because of lack of data. As for rating prediction, it can be embedded in the now broad literature of classification problems \cite{friedman2001elements}. The idea of employing machine learning techniques in credit rating scoring has been explored before \cite{altman1994corporate}, but in that case the predictors for the rating are all derived from balance sheets, so the results are not comparable with ours. Other works use more heterogeneous information to predict the rating \cite{WILSON1994545, Grunert2005509, Lee200767, parnes2012approximating}.

This paper contributes to these streams of literature in several aspects. First, we investigate the topological properties of payment networks by considering standard network metrics, such as degree and strength distribution and components decomposition.
We find that the large payment networks investigated in this paper share the properties observed in other complex networks, namely they are sparse but almost entirely made of a single component, they are scale free and small world.  Then, we look into the distribution of risk of firms in the network of payments in order to quantify the dependence between the network property of a node or a group of nodes and the risk of the firm represented by the node(s). The main and most innovative contribution of this paper is to document the existence of such correlations. We find an homophily of risk, i.e. the tendency of a firm to interact with firms with similar risk. This is a two nodes properties, but a similar behaviour is observed, even more clearly, also at larger aggregation scales. Communities of firms, detected by using different methods, often display a statistically significant abundance of firms of a specific risk class, indicating the tendency of firms with similar rating to be linked together through payments. Risk is therefore not spread uniformly on the network, but rather it is concentrated in specific \emph{areas}. This implies that an idiosyncratic shock on a single firm can propagate more or less quickly depending on the local network structure and the community the node belongs to. The last contribution, is to exploit this correlation between risk of a firm and network characteristics of the corresponding node to predict the risk rating of the firm using network properties \emph{alone}. To this end, we employ machine learning techniques to build classifiers for risk rating whose inputs are only network properties (e.g. degree, community, etc.). We show that our classification method has a good performance both in terms of accuracy and of recall and that outperforms significantly random assignments. 

\section{The network of payments}\label{sec:network}
\subsection{The dataset}
The investigated dataset contains information on payments between more than two million Italian firms and is built from transactional data of the payment platform of a major European bank. Transactions are registered with daily granularity for the year 2014, for a total of $47$M records, each of which includes the two counterparts involved, date, type, amount and number of transactions in the same day. Transactions are originally identified by account, but in case of customers and former customers, multiple accounts associated to the same firm are aggregated into a single entity\footnote{In order to comply with privacy regulation any payment from or to physical persons is excluded, moreover the filter is implemented to exclude any ambiguous record.}. This results in a total of $2.4$M entities (which will be referred to as firms, for brevity) operating through the platform during the whole investigated period.
The counterparts can be of different types. In principle, any firm or public body can make use of the platform, but in practice in most cases at least one is a customer of the bank. Similar considerations hold for the total amount exchanged: in each month more than $50\%$ of the volume is transferred between customers, and it rises to above $95\%$ when considering transaction with at least one customer involved. More details on the dataset and some descriptive statistics is presented in \ref{a1:network}. For customers, the dataset contains information on the economic sector and on the internal rating of the firm on a three value scale: Low (L), Medium (M), and High (H) risk.
 
\subsection{Networks definition and basic metrics}
A network, or graph, is identified by two sets: $V$, the sets of nodes with cardinality $\abs{V}=n$,  and $E$, the sets of links or edges, with cardinality $\abs{E}=m$. The latter is the collection of \emph{ordered} pairs of connected nodes. In our case, we also take into account the strength of interactions so a weight $w_{ij}$ is associated with each link. Starting from transaction data, payment networks are constructed as follows: given a time window, each node represents a firm active in that period, if there is payment between two firms a link from the source to the recipient is added, with weight equal to the payment amount. If multiple transactions occur between the same (ordered) pair of nodes, the weight of the link is the sum of the amounts of the payments. Therefore for each time period we construct a directed and weighted network. The time window of analysis may vary depending on the type of information one wants to extract from the dataset.  In the following the focus will be on monthly networks, for which results are quite stable, at the cost of dealing with fewer and larger graphs. For the period covered by the dataset, each monthly network consists on average of $n=1$M nodes and $m=3.2$M links with the lowest activity in August and the highest in July (see \ref{a:dataset}). The density $\rho=\frac{m}{n(n-1)}$ is thus small, resulting in a so called sparse network. Nevertheless this low density does not imply a disaggregated system. Indeed for all the networks the diameter is very small compared to the size: on average starting from a node, one has to pass at most 19 links to reach any other node in the weakly connected component (see Table \ref{tab:basics}). Thus the networks have the so called small-world property. 

\begin{table}[h]
\centering
\title{}
\caption{Basic metrics of the network of payments}\label{tab:basics}
\begin{tabular}{c|cccccc}
&nodes&links&	in-degree&out-degree &	density&diameter\\
month&$n$&$m$&	$\mathbb{E}[k^{(in)}]$&$\mathbb{E}[k^{(out)}]$ &	$\rho=\frac{m}{n(n-1)}$& $d$\\
\hline
Jan	&	1,000,555	&	3,271,861	&	6.61	&	4.20	&	$3.27\cdot 10^{-6}$	&	20\\
Feb	&	997,006	&	3,067,029	&	6.11	&	3.98		&	$3.09\cdot 10^{-6}$	&	18\\
Mar	&	1,018,164	&	3,146,559	&	6.19	&	3.98	&	$3.04\cdot 10^{-6}$	&	18\\
Apr	&	1,047,706	&	3,346,763	&	6.52	&	4.09	&	$3.05\cdot 10^{-6 }$&	19\\
May	&	1,048,803	&	3,359,315	&	6.58	&	4.08		&	$3.05\cdot 10^{-6	}$&	20\\
Jun	&	1,039,876	&	3,239,886	&	6.30	&	4.02		&	$3.00\cdot 10^{-6}$&19\\
Jul	&	1,091,393	&	3,510,435	&	6.44	&	4.14		&	$2.95\cdot 10^{-6}$&20\\
Aug	&	891,587	&	2,319,697	&	5.21	&	3.44	&	$2.92\cdot 10^{-6}$	&	19	\\
Sep	&	1,041,124	&	3,465,233	&	6.80	&	4.25		&	$3.20\cdot 10^{-6}$&20\\
Oct	&	1,066,044	&	3,289,946	&	6.11	&	4.00	&	$ 2.89\cdot 10^{-6}$&18\\
Nov	&	1,023,692	&	3,103,365	&	6.15	&	3.90	&	$ 2.96\cdot 10^{-6}$&18\\
Dec	&	1,052,975	&	3,000,284	&	5.60	&	3.74	&	$2.71\cdot 10^{-6}$	&19\\
\end{tabular}

\end{table}

\subsection{Networks topology}
When considering a small number of firms, one would expect simple topologies: one firms is the supplier of intermediate products for another firm, resulting in a line  (the simplest supply chain), or one firm is a supplier or a buyer for  many others firms, resulting in a star network. Instead what is observed is a much more complex organisation, with a non negligible presence of cycles.

At a very coarse level, it is possible to identify two large classes of firms.
The first constitute the core of the network, which includes approximately $20\%$ of the nodes and  more than half of the links. This core has a density an order of magnitude larger than that of the whole network and it is characterised by the fact that any pair of firms is connected, directly or via intermediaries. Around $60\%$ of the total volume circulates among the nodes of the core. 
The other class is made  of payers-only, i.e nodes that have no incoming links. These represent each month about one half of the active firms and their activity is sporadic. To better understand the role of this significant subset of firms we check their customer status and we find that the majority of them are unclassified, and that their number is larger than one expects from the unconditional distribution among all the firms. This means that likely they are not customers and, more importantly, almost no information, for example about risk, is available on them. For further details on this refer to \ref{a:further}.

We now turn our attention to the distribution of degree and strength. In our case the in- (out-) degree is the number of payers (payees) of a given firm and the corresponding amount of Euro. For the monthly aggregation case the average in- and out-degree of a firm is $6$ and $4$, respectively (see Table \ref{tab:basics}). These low values are a direct consequence of the low density of the network. However the degrees and the strengths are extremely heterogeneous as testified by the degree and strength distribution. 

\begin{figure}[h]
\centering
\includegraphics[scale=0.35]{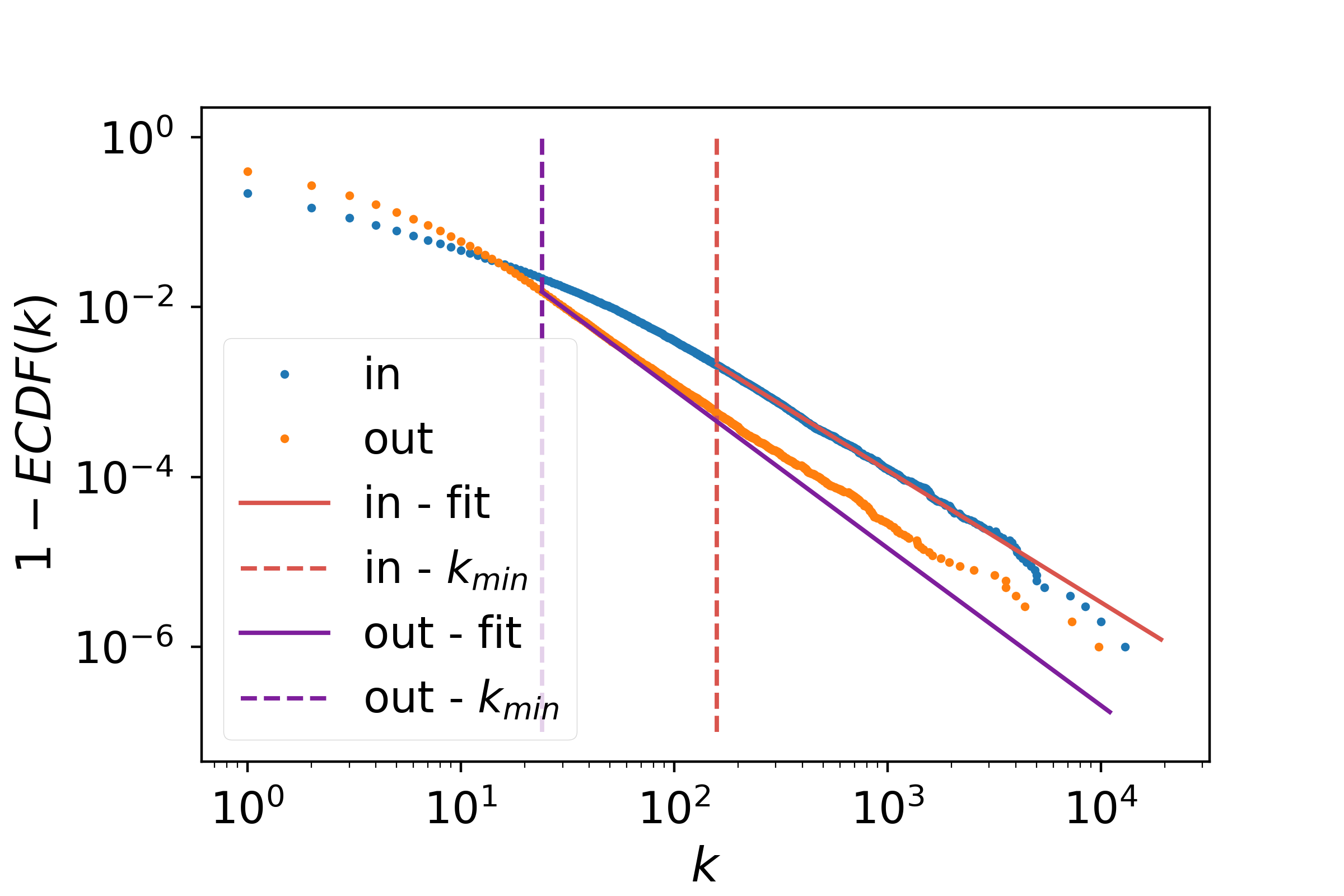}
\includegraphics[scale=0.35]{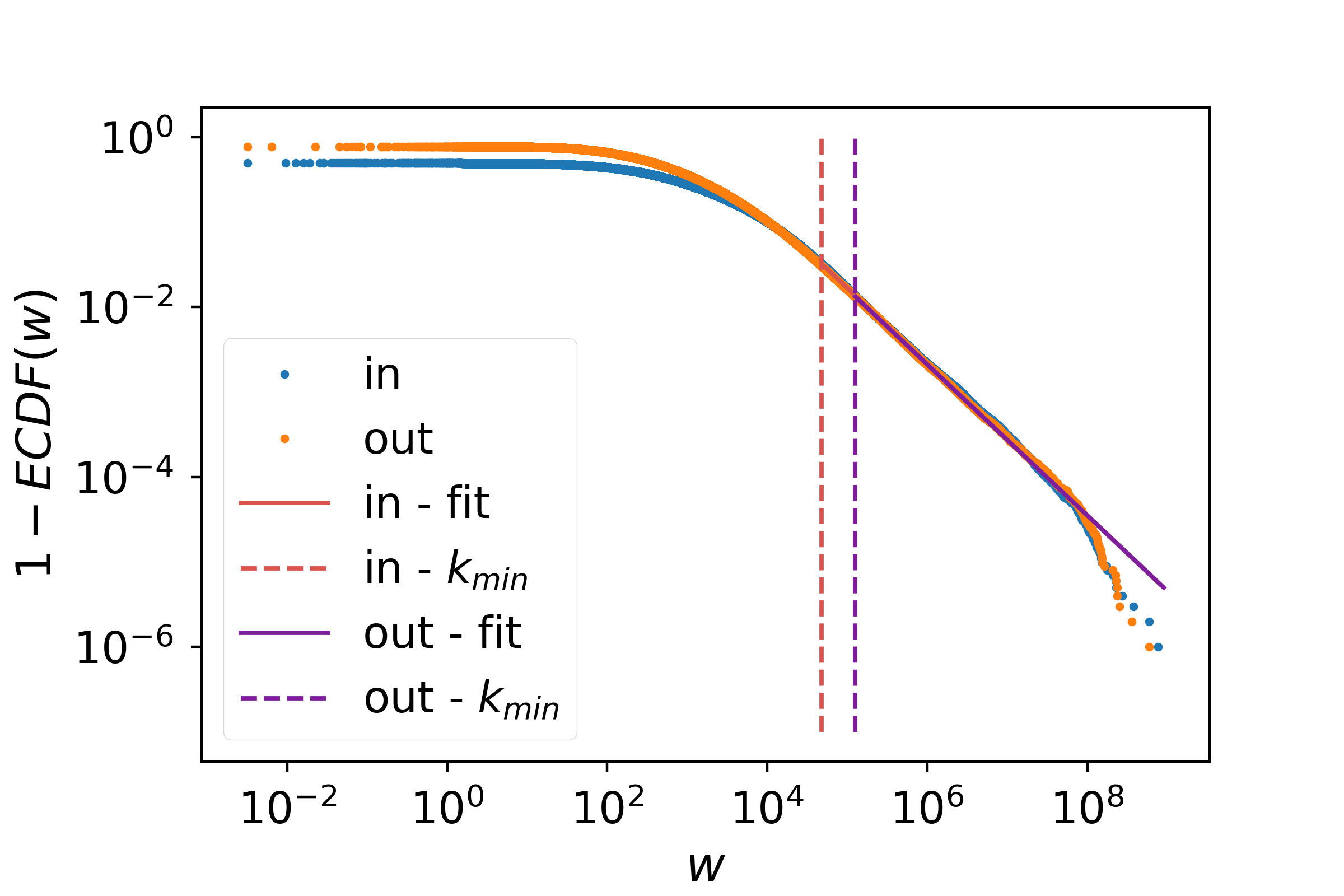}
\caption{Empirical complementary cumulative degree (left) and strength (right) distributions and their power law fit. The scale is logarithmic for both axes. Data refers to January, but results are similar for the other months.}\label{fig:pl-str}
\end{figure}

Figure \ref{fig:pl-str} shows the empirical cumulative distribution for these two quantities in a double logarithmic scale. The approximately straight line indicates the presence of a fat tail with a power law behaviour. The fit of the exponent supports the observation that in- and out- degree distribution data are consistent with a power-law tail and the estimated exponents are around  $2.6$ and $2.8$, respectively. Similarly, in-strength and out-strength are well fitted by power-law distributions of exponents around $2.1$ and $2$, respectively. 
Despite the fact that a large fraction of nodes is different in each month, the tail exponents are remarkably stable (see Table \ref{tab:pl-deg} of the \ref{a:further}). 

This scale free behaviour is quite ubiquitous in complex networks has been found in many other real economic and financial networks \cite{serrano2003topology,garlaschelli2005structure, boginski2005statistical,boss2004network,kim2002weighted,huang2009network,takaaki2009}. 
The fat-tailed distribution for the degree has two interesting consequences: first, there is no characteristic scale for the average degree or strength; second, there are a few nodes that act as hubs for the system, in the sense that, having a large amount of connections, many pairs of nodes are connected through them. This partially explains the low values for the diameter.

Finally, we measure the tendency of firms to be connected to firms which are similar with respect to some attribute, namely the number and the total volume of connections (i.e. degree and strength). Following \cite{newman2002assortative}
, we compute the assortativity coefficient for a categorical variable, 
\begin{equation}
r=\frac{\sum_i e_{ii}-a_ib_i}{1-\sum_i a_ib_i}\label{eq:assort_coeff}
\end{equation}
where $e_{ij}$ is the fraction of edges connecting vertices of type $i$ and $j$, $a_i = \sum_ j e_{ij}$ and $b_j = \sum_i e_{ij}$. It is $r_{\text{max}} = 1$ for perfect mixing, while when the network is perfectly disassortative (each node connects to a node of a different type) it is
$r_{\text{min}}=-\frac{\sum_ia_ib_i}{1-\sum_i a_ib_i}$. Using the number of connections as categorical variable, an high value for the assortativity coefficient indicates that highly connected firms tend to interact significantly more than average with other highly connected firms. Similar reasoning holds using the volume exchanged as categorical variable.

Beside the entire graph, we also consider the subgraph of firms with rating and the subgraph of customers. The assortativity coefficient is consistently slightly negative for both attributes, for all  months and graphs, namely around $-0.03$ for the entire graph and the subgraph of firms with rating, and $-0.04$ for the subgraph of customers, with no strong differences among months and attributes. Table \ref{tab:assort} of \ref{a1:network} reports the summary of values of the assortativity coefficient for each month. 
A possible explanation can be that large, very interconnected firms are connected to many subsidiaries which in turn do not engage with many other firms, being their business almost exclusively focussed on the relationship with the large and central firms. 

To summarise, each month the payment network of firms is very sparse but almost entirely connected. Half of the firms appear in the network as payers only (no incoming links) and they are mainly unclassified with respect to customer status, so no much information is available on them. Of the remaining nodes, almost half constitutes the denser core of the network where more than a half of the transactions occur and above $60\%$ of the volume circulates. Finally the network is small world, scale free, and slightly disassortative both for degree and for strength.
\section{Risk distribution and network topology
}\label{sec:risk}

In this Section we investigate the distribution of risk of firms in the network of payments. We are interested in measuring the dependence between the network property of a node or a group of nodes and the risk of the firm represented by the node(s). 
We proceed in a bottom-up fashion, zooming out from single nodes to subsets. 
At first we consider a firm's local property (the number of connections) and we check if it correlates with the risk. Then we consider pairs of linked firms and measure the homophily in risk, i.e. whether firms with similar risk profile tend to do business together and thus to be linked. Finally, we divide firms into subsets induced by the network structure and we check whether the inferred subsets are informative with respect to the riskiness of the composing firms. 
Specifically, we partition the network in groups (or communities) of firms by using only network information, and we test if the distribution of risk within each group is statistically different from the global one. Thus the goal is to understand if the inferred communities are homogeneous with respect to the risk profile of the composing firms: a community with many firms with high risk rating is a clear indication of financial fragility and a possible source of instability, since the distress of one or few firms of the community is likely to propagate to the other firms. 

For the sake of brevity, in the following the analysis is carried on for one month, but results are consistent for all the months.

\subsection{Degree and risk}

\begin{figure}[htb]
\centering
\includegraphics[scale=0.6]{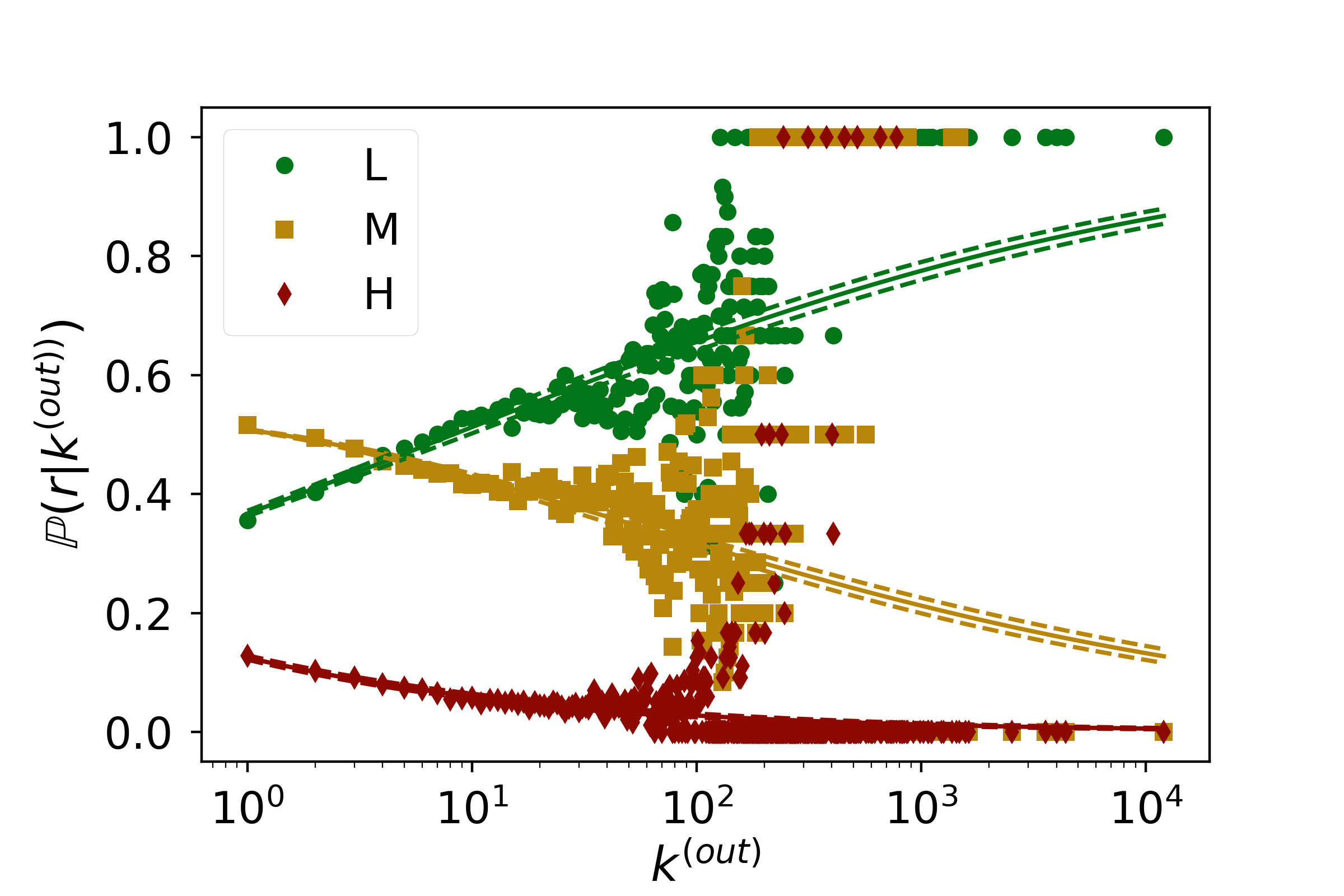}
\caption{Probability of rating of a firm conditional to its out-degree. The solid lines show the fitted multinomial logistic distribution, with its confidence intervals (dashed lines) in matching colours.}\label{fig:m01_degdist}
\end{figure}

The first investigation is on the relation between the degree of a firm and its risk. 
The probability for each risk level $r\in{L,M,H}$ conditional to the out-degree is computed\footnote{The results for the in-degree are qualitatively very similar.} and plotted against the degree. The results are shown in Figure \ref{fig:m01_degdist}.
We notice an interesting correlation between degree and risk: small degree nodes are more likely medium risk firms, whereas large degree nodes are more likely low risk firms. The high risk firms are more evenly spread across degrees, even if a larger fraction is observed for low degree nodes.
To assess if the three curves are statistically different we perform a multinomial logistic regression on data \cite{greene2003econometric} (the solid lines in the plot). This choice is justified by the fact the quantities just described are the probabilities of outcomes in a multi-class problem  given an independent variable (the degree). The estimated probabilities follow quite closely the trend of the empirical distribution and the coefficients are all significant. More detailed results of the fit are given in table \ref{tab:logit_results} of \ref{a:deg_risk} (first two columns).

The correlation just highlighted can be, at least in part, influenced by the effect of the size  of the firm (in term of assets value from the balance sheet): a large firm is usually considered less risky than a small one, at the same time, a larger size generally implies a higher number of connections, as seen for example in the interbank network \cite{bargigli2015multiplex}. As the size of firms is not available to us, we use the sum of the incoming and outgoing amounts as proxy. Defined in this way, the size has a Pearson correlation of around $0.19\,,0.15$ with in- and out- degree, but a Spearman rank correlation of $0.67\,,0.57$ respectively. To control for the effect of the size, we repeat the same procedure on subsets of firms, grouping according to their size into tertiles. The results, not shown here but available upon request, hold true also on the sub-samples, but slightly less sharply. We repeat the multinomial logistic regression adding the size tertiles among the predictors, and we still obtain statistically significant coefficients (last four column in \ref{tab:logit_results} of \ref{a:deg_risk}).

Similarly, the three conditional degree distributions given the rating  result statistically different, as for every month all pairs reject the null hypothesis in the 2-sample Kolmogorov-Smirnov test \cite{smirnov1939estimation}.
Therefore topological characteristics (the degree) of the node can be used to obtain information on the riskiness of the corresponding firm. From a risk management perspective this is an important results, since on average highly connected nodes are also less risky.

\subsection{Assortative mixing of risk}\label{s:assort}
The next step is to check whether risk is correlated with direct connection preferences. To clarify this point, we consider two features: the assortative mixing of the risk and the conditional distribution of rating given the distance. 

In the first case we compute a  weighted variant of  the assortativity coefficient in Eq. \eqref{eq:assort_coeff} using as categorical variable the risk rating rather than the degree or the strength. In practice, the quantities $e_{ij}$ are substituted by $\tilde e_{ij}$, the fraction of volume from nodes of type $i$ to nodes of type $j$. The reason for this choice is to mitigate the impact of the aforementioned large number of uncategorised payers. In most cases their links are associated with low volume and few transactions. Also, customer firms, even if they represent only around $1/3$ of the firms, exhibit a generally more intense activity, both in terms of number of transactions and of volume, hence accounting for the stronger ties between the firms.

The metric is positive for all the three graphs, $0.070$, $0.157$, $0.163$ for the whole set, the nodes with rating, and the customers, respectively, with significant variability across the months but always positive sign\footnote{Results for the standard assortativity coefficient are quite different, and the choice of the subgraph appears to be crucial. When considering the entire network, the assortativity coefficient is negative, around $-0.07$, hence indicating a slightly disassortative behaviour with respect to risk. The subgraphs, instead, show an assortative tendency, with coefficients around $0.025$ and $0.038$ for the nodes with rating and for customers, respectively. This shift can be explained again by the impact of the large number of uncategorised nodes.}.
In Table \ref{tab:assort_risk} of \ref{a:assort}, the summary of values of the assortativity coefficients for each month is presented.

With the same quantities $\tilde{e}_{ij}$ we define metrics to assess different preferences in connection between incoming and outgoing payments. We test if firms are more concerned with the risk of payers than of the payees by testing for different risk distribution between incoming and outgoing connection.  To discriminate between these two cases, for each node $i$ we compute the percentage excess of volume with respect to the average toward nodes in certain risk class and we group according the rating of the node. The distributions are compared using Mann–Whitney U test \cite{mann1947test}. This non-parametric test allow to assess if one distribution is  stochastically greater than the other. Details on the metrics and the test performed are given in \ref{a:assort}. We find that it is likely that firms are, at least in part, aware of the riskiness of their counterparts and results suggest they use this information in choosing their business partners. However the hypothesis that incoming payments show a more marked preference for low risk is not supported by data. Moreover the overall positive assortativity is mainly due to low risk nodes. This suggests that low risk firms are more careful in the choice of their business counterparts, possibly also because their relative larger creditworthiness allow them to find available partners more easily.    

The quantities considered so far in this section are pairwise comparisons between the rating of nearest neighbours, and give an aggregate measure. A possible\footnote{An alternative strategy to go beyond first order neighbours in the computation of assortativity has been recently proposed by \cite{arcagni2017higher}.} way to enrich this information is to consider the conditional distribution of rating for nodes at a given distance\footnote{The distance between nodes in a network is defined as the length of the shortest directed path connecting two nodes, where a path is a sequence of links. Clearly, in a directed network in general $d(u,v)\ne d(u,v)$ and moreover $d(u,v)$ can be not defined (or $\infty$) if there is no path from $u$ to $v$.}
and to compare it to the unconditional distribution. In the case of no influence of the rating on the connection pattern, the conditional distribution of risk given the distance should be statistically undistinguishable from the null unconditional distribution. To test if this is the case, we first compute the distance between all the nodes for which the rating is available. Then for any fixed $k$, the occurrences of ratings are computed by looking at the set of pairs at distance $k$. Finally, the estimated distributions are tested against the null one with an hyper-geometric test, as explained in details in \ref{a:test}.

\begin{figure}[htb]
\centering
\includegraphics[scale=0.6]{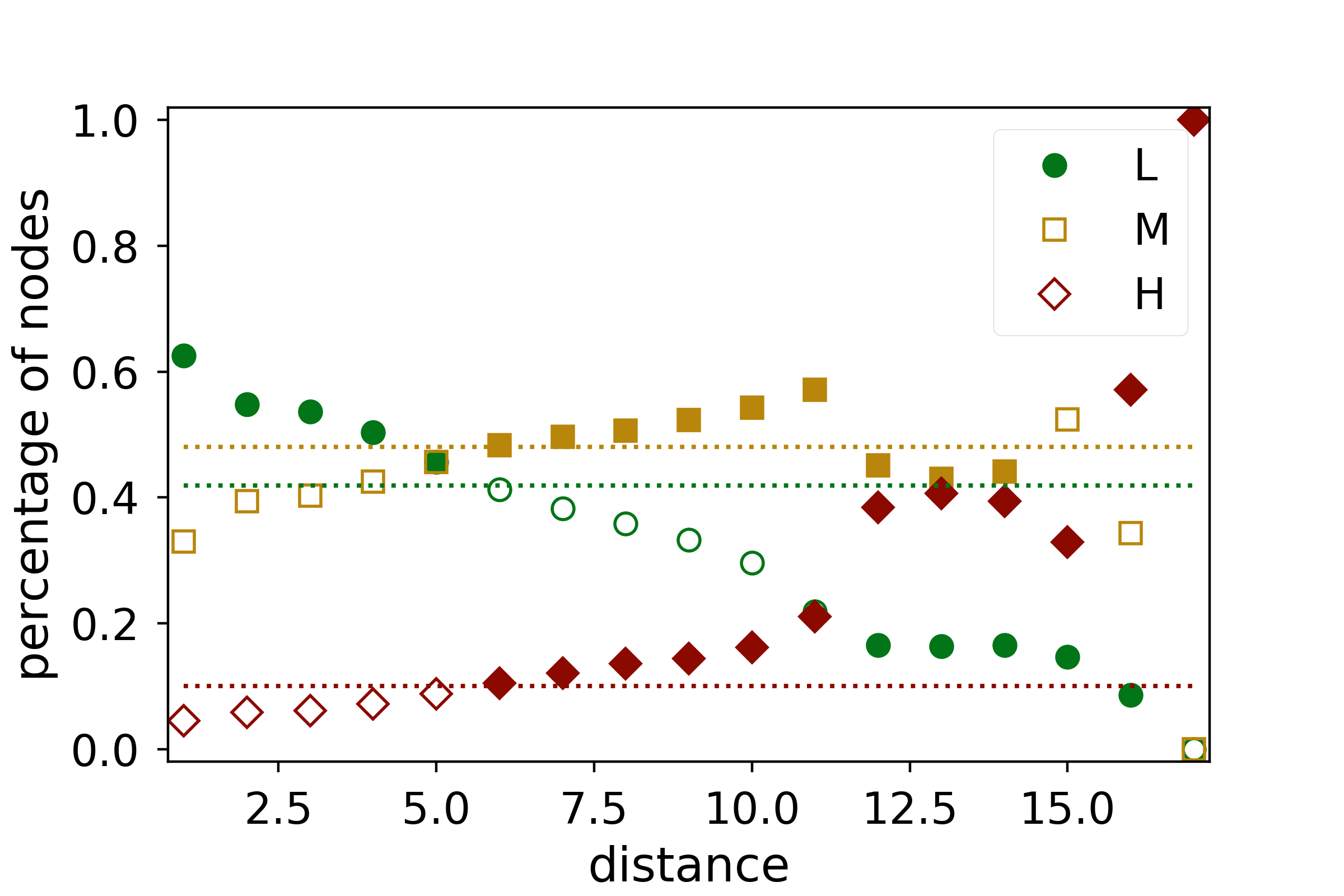}
\caption{Distribution of ratings for nodes at distance $k$ from a node with rating L. The dashed lines are the unconditional (null) distribution of ratings among nodes in the entire sample. A full marker indicates that the over or under representation with respect to the null distribution is statistically significant in the hyper-geometric test at $1\%$ significance level with Bonferroni correction.
}\label{fig:m04_degdist}
\end{figure}

Results for April are summarised in Figure \ref{fig:m04_degdist}, which considers the case when the source node is in class L. Results are similar when considering a medium or high risk source. For each $k$ a marker indicates the percentage of nodes with low (green circles), medium (yellow squares) or high (red diamonds) risk at distance $k$. A marker is full when the percentage is statistically different from the null distribution (the dashed lines, with matching colours).

We note that up to distance 5 the class of low risk firms is significantly over-represented in the distributions. At greater distances, medium  and high risk groups are over-represented.  This means that more steps in the networks are necessary to reach riskier firms. This fact is particularly interesting when considering that each firm is in theory unaware of others firms' ratings and in some cases even its own. 

When considering the same quantities for incoming paths, results (not shown) are very similar, namely at short distances the low risk class is over-represented, while medium and high risk nodes are over-represented for longer distances. 

A possible explanation for these observations is that among the hubs of the systems (i.e the most connected nodes) firms with rating L (i.e the most creditworthy) constitute the vast majority. This holds true when considering both in-coming and out-going links, and including also the nodes with no rating. Moreover, they are in the denser core previously described, while many high risk firms have a few or no out-going links and they are peripheral in network. This asymmetry in the \emph{position} in the network is observed also when considering the distribution of the closeness centrality \cite{newman2010networks} of the nodes, i.e the harmonic mean of the distances to all other nodes, conditioning on the risk class (not shown\footnote{Results available upon requests}).

\subsection{Network organization and risk}
In this Section we study the relation between the organization of the network at a more aggregate level and the distribution of risk. We are interested in two types of organization of networks into groups. The first is the {\it modular} organization: each module is composed by nodes, which are much more connected among themselves than with the rest of the network. In economic terms modules could represent, for example, firms operating in the same region or area, and the high density of the module reflects the fact that payments are more frequent with geographically close firms. We saw before that the network shows an assortative tendency with respect to risk, so we want to test if the homophily on risk can be observed beyond pairwise relationship.

The second is a {\it hierarchical} organization. Since the payment network is directed, we look for a ranked partition (i.e. each group of nodes is labelled with an integer from $1$ to the number of groups $M$) such that most links are from nodes of low rank classes to nodes in high rank classes. This type of organization could represent, for example, a supply chain and the flow of payments between the firms of a group and those in the group in the next rank class reflects the (opposite) flow of goods or services. This classification is important because a high risk concentration in low class nodes of a strongly hierarchical network can trigger a cascade of distress in the higher rank classes.

Modularity and hierarchy are conceptually opposite as the first penalises connections towards other groups, which instead are encouraged in the latter (provided that they go from low rank to high rank nodes).

For each metric, we proceed in the following way: 
\begin{enumerate}[noitemsep,label=\roman*.]
\item find the optimal partition according to the criterion;
\item compute the distribution of ratings within each subset of the partition;
\item test whether such \emph{local} distribution is statistically different from the overall distribution of ratings by employing the hypergeometric test used in the previous Section and described in \ref{a:test}. In order to have a sample large enough to perform the test, we only consider subsets with at least 500 known ratings.
\end{enumerate}

We showed so far that the structure of the payments network is very complex. Since our goal is to obtain information on the risk of the firms, it can be helpful to filter the network before performing communities detection, in order to keep the most relevant connections. Thus we focus on the subgraph of customers. The reasons for this choice are many. First, the percentage of nodes with rating active every month is quite low, around $20\%$, but it raises to $70\%$ when considering only the customers (see Table \ref{tab:customer_rating} in \ref{a:dataset} for a summary). This will help having a more informative local distribution of risk when considering subsets of nodes.  Secondly, more than a half of the volume is transferred between customers (see Table \ref{tab:volume-by-status} in \ref{a:dataset}), so even if a large fraction of transactions is dropped, we are mostly pruning weak connections, while keeping the strongest ones. Finally, as it has been shown in the previous Subsection about assortativity, considering the entire network can be misleading, especially when looking at the connections without considering the weights, as it will be necessary for some metrics.

\begin{figure}[p]
\setlength\figureheight{0.65\columnwidth} 
\setlength\figurewidth{\columnwidth} 
\centering
\definecolor{l}{rgb}{0.00392156862745098, 0.4588235294117647, 0.09019607843137255}%
\definecolor{m}{rgb}{0.7215686274509804, 0.5254901960784314, 0.043137254901960784}%
\definecolor{h}{rgb}{0.5490196078431373, 0.03529411764705882, 0.0}%
 
\begin{tikzpicture}

\begin{axis}[%
width=\figurewidth,
height=\figureheight,
axis line style={draw=none}, 
xtick=\empty, 
ytick=\empty,
name=mod,
legend entries={L,M,H},
legend style={draw=none,at={(0.95,0.5)},anchor=east},
]

]
\addplot [forget plot] graphics [xmin=0,xmax=16,ymin=0,ymax=0.6] {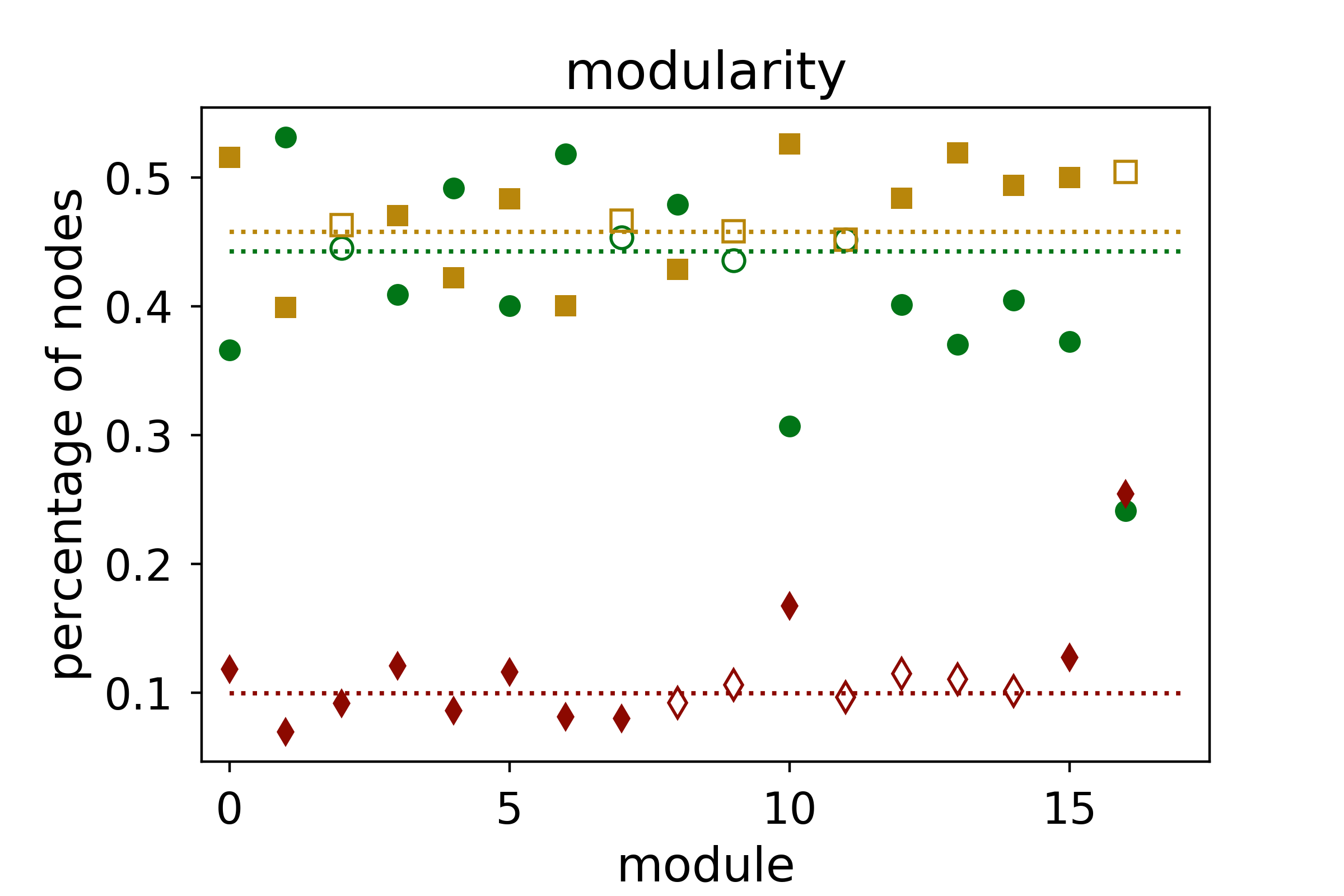};
\addlegendimage{only marks, mark=*,color= l}
\addlegendimage{only marks, mark=square*,color = m}
\addlegendimage{only marks, mark=diamond*,color= h}
\end{axis}

\begin{axis}[%
width=\figurewidth,
height=\figureheight,
axis line style={draw=none}, 
xtick=\empty, 
ytick=\empty,
name=hier,
at =(mod.below south east),
anchor= above north east,
legend entries={L,M,H},
legend style={draw=none,at={(0.95,0.5)},anchor=east},
]

]
\addplot [forget plot] graphics [xmin=0,xmax=16,ymin=0,ymax=0.6] {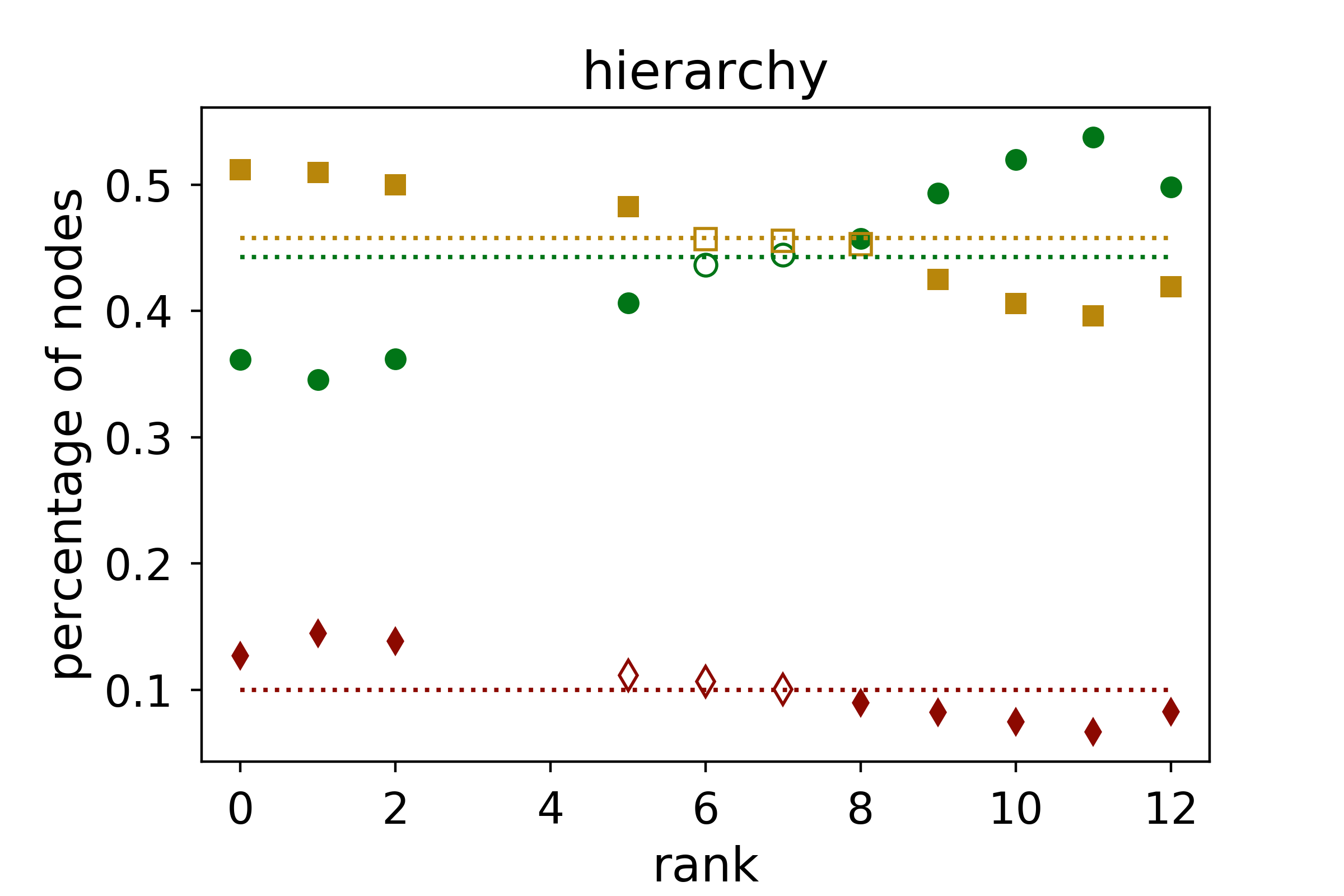};
\addlegendimage{only marks, mark=*,color= l}
\addlegendimage{only marks, mark=square*,color = m}
\addlegendimage{only marks, mark=diamond*,color= h}
\end{axis}

\end{tikzpicture}%
\caption{Distribution of ratings in the three partitions, modularity (top), hierarchy (bottom). The dashed lines are the unconditional (null) distribution of ratings among nodes in the entire sample. A full marker indicates that the over (above the dashed line) or under (below the dashed line) representation with respect to the null distribution is statistically significant in the hypergeometric test at $1\%$ significance level with Bonferroni correction. 
}
\label{fig:m01_classes}\end{figure}

\subsubsection{Modular structure}

One of the standard methods for inferring a modular structure in a network is via modularity maximization. This method divides nodes into subsets, called modules, such that nodes are well connected with other nodes in the same module and there is a smaller number of links with nodes in other modules. Given a partition $P$ in modules $C$ the modularity is
\begin{equation}
Q=\frac{1}{2m}\sum_{C\in P}\sum_{i,j\in C}\left(A_{ij}-\frac{k^{in}_ik^{out}_j}{2m}\right)
\end{equation}
where $A_{ij}$ is the $(i,j)$ element of the adjacency matrix and $k^{in}_i$ ($k^{out}_i$) is the in- (out-) degree of node $i$. The optimal partition is the one which maximizes modularity. Despite the associated optimisation problem is NP-Hard, fast and reliable heuristics for an approximate solution exist, and here the well known Louvain method \cite{blondel2008fast} is employed.

In each month we find that the optimal partition has around $2,000$ modules. These are really heterogeneous in size: for example, the 13 largest ones cover more than $95\%$ of the nodes of the network. We perform the hypergeometric test of the null hypothesis of an homogeneous distribution of risk in each module with at least 500 known ratings. These are less than $1\%$, around 19 every month. 
(see Table \ref{tab:test_mod_sum} in \ref{a:test} for more details).
These are clearly very large modules but a significant number of them shows an over or under-expression of one or two risk classes. 

For some specific module it is possible to draw statistical robust conclusions on its risk profile. The top panel of Fig. \ref{fig:m01_classes} shows the over- or under-representation for the largest modules for January. The seventh module, for example, has an over-representation of firms with low risk and an under-representation of the other two risk profiles, thus it represents a group of firms with small risk. On the contrary the eighth module has an over-representation of highly risky firms and under-representation of low risk firms, representing a possible warning for the bank. 

\subsubsection{Hierarchical organization}

We now consider explicitly the directed nature of the payment graph and the hierarchical organization of the network. An ordered partition is such that each subset is associated with an integer number (rank) $r\in \{1,...,M\}$. A graph has a hierarchical organization if nodes are more likely linked to other nodes with a higher rank \cite{simon1991architecture},  such as in military organizations or in administrative staff. Finding the optimal ordered partition and revealing the hierarchy of a graph is in general complex and requires the minimization of a suitable cost function, similarly to what is done with modularity. 

In this paper we use a recently proposed cost function \cite{Gupte2011}. Given a rank function $r:V\to \{1,...,M\}$, the cost function penalizes links from a high rank node to a low rank node. The penalization is a linear function of the difference between the ranks. Thus the optimal hierarchical partition is obtained by solving the optimisation problem 
\[A^*=\min_{r\in \mathcal{R}}\sum_{(u,v)\in E} f(r(u)-r(v))\,,\]
where $\mathcal{R}$ denotes the set of all ordered partitions and the cost function is 
\[f(x)=\begin{cases}x+1& x\geq0\\
0&x<0\end{cases}\,.
\]
The \emph{hierarchy} of the graph is defined by
\[
h^*(G)=1-\frac{A^*}{m}\,.
\]
By definition, $h\in[0,1]$, and 0 is the value for the trivial partition with only one set, while $h=1$ is obtained when the network is a Directed Acyclical Graph and it signals a perfect hierarchy. The linear choice of the penalization function is convenient because the associated optimisation is solvable in polynomial time and few exact algorithms exist  \cite{Gupte2011,tatti2017tiers}, while non-linear forms can lead to NP-hard problem.

We apply the hierarchy detection to the monthly networks of payments and the results are summarized in Table \ref{tab:test_ag_sum} of \ref{a:test}. First of all we notice that the number of inferred classes, roughly $18$, is much lower than in the modular case. Moreover the size of the classes is much more homogeneous. The value of $h$ is also quite stable, around $0.75$, indicating a strong hierarchical structure, a remarkable result considering that we are studying only the customers network.

We now consider the distribution of risk in each class and we study the over or under-expression of certain levels of risk as a function of the rank of the class in the inferred hierarchy. The test rejects the null hypothesis of uniform risk distribution a considerable number of times (also compared with the total number of subsets in the partitions). As displayed in the bottom panel of Figure \ref{fig:m01_classes}, low rank classes have an over-expression of high and medium risk firms, while middle and low rank classes (i.e. $r\in [8,12]$) have an over expression of low risk firms and an under-expression of medium and high risk firms. More details on the test results are  given in This empirical evidence may signal the presence of paths of risk propagation, since low rank firms, typically more risky, are payers of high rank firms, which are instead less risky.  
We now consider the distribution of risk in each class and we study the over or under-expression of certain levels of risk as a function of the rank of the class in the inferred hierarchy. The test rejects the null hypothesis of uniform risk distribution a considerable number of times (also compared with the total number of subsets in the partitions). As displayed in the bottom panel of Figure \ref{fig:m01_classes}, low rank classes have an over-expression of high and medium risk firms, while middle and low rank classes (i.e. $r\in [8,12]$) have an over expression of low risk firms and an under-expression of medium and high risk firms. More details on the test results are  given in This empirical evidence may signal the presence of paths of risk propagation, since low rank firms, typically more risky, are payers of high rank firms, which are instead less risky.  
We now consider the distribution of risk in each class and we study the over or under-expression of certain levels of risk as a function of the rank of the class in the inferred hierarchy. The test rejects the null hypothesis of uniform risk distribution a considerable number of times (also compared with the total number of subsets in the partitions). As displayed in the bottom panel of Figure \ref{fig:m01_classes}, low rank classes have an over-expression of high and medium risk firms, while middle and low rank classes (i.e. $r\in [8,12]$) have an over expression of low risk firms and an under-expression of medium and high risk firms. More details on the test results are  given in \ref{tab:test_ag_sum} in \ref{a:test}. This empirical evidence may signal the presence of paths of risk propagation, since low rank firms, typically more risky, are payers of high rank firms, which are instead less risky.  

\subsection{Discussion}
Both investigated partitions give interesting insights on the relationship between risk and network structure. On one side, the percentage of rejected tests in the case of modularity partition is consistent with the observed assortativity of risk. It may be noticed that the preference for low risk business partners is not always a realistic option, because in some sectors business partners are not replaceable for geographical reasons. To better assess this point, one possibility could be to  include the comparison between modules and geographical location of firms, which is not available to us. On the other side, the hierarchical partition appears to follow the risk distribution slightly better and this is probably related to the peculiar conditional distribution of risk with respect to the distance described in Subsection \ref{s:assort}. Indeed, given the fact the high risk nodes are over represented for longer distances, they should be located in extreme positions in the ranking, either at the top or at the bottom, and this is what is observed. 
It must be stressed that in the case of the two methods chosen here, one does not exclude the other, as they give different and complementary standpoints for interpretation.  In this sense a \emph{multi-dimensional} perspective is needed, where the dimensions are the mechanisms that either favour or discourage the creation of business relationships.
\section{Missing rating prediction using payments network data}\label{sec:prediction}
In the previous Sections we showed that network metrics can be informative of the risk of a firm. It is therefore natural to ask whether it is possible to predict the missing risk rating of a firm by using \emph{only} information on network characteristics of the corresponding node, as well as risk rating of the neighbour firms. This problem is particularly relevant since we noticed that around $30\%$ of the customers in the dataset do not have a rating and this percentage is even higher when the entire dataset is considered (see table \ref{tab:customer_rating} in \ref{a:dataset}). 

Here we use network characteristics as predictors for the missing ratings into well known methods of machine learning for classification problem. 
The predictors we employ are the following:
\begin{enumerate}[noitemsep,label=\roman*.]
\item in- and out-degree;
\item weighted fraction of (in- and out-) neighbours with a given rating (H,M,L or NA)
\item rank of the class in the hierarchy inferred by agony minimisation;
\item membership in community inferred by modularity maximisation;  
\item sum of in- and out-strength.
\end{enumerate}
The fractions in (ii.) are computed considering the amount (weight) of each payment and are together a measure for rating assortativity, while (v.) is a proxy for the size.
Data are preprocessed following \cite{friedman2001elements} so that variables are comparable in order of magnitude, as detailed in Appendix \ref{a:preparation}. These transformations result into a total of 25 predictors.
The dataset is the one which includes only the customers, and we consider the monthly network for January. In order to assess the performance of the prediction,  we train each model using $75\%$ of the data, and the remaining $25\%$ is used for testing.

We consider three methods for classification:
\begin{enumerate}[noitemsep,label=\roman*.]
\item multinomial logistic; 
\item classification trees; 
\item neural networks. 
\end{enumerate}
See \cite{friedman2001elements} for a review of these methods.

The class \emph{H} is under-represented in the sample, as it includes only around $10\%$ of the firms with rating. This affects the ability of any classifiers to recover this class. This is undesirable, since the class $H$ the most critical for the riskiness.

To address this issue we proceed with a 2-step classification strategy for all the three methods. The intuition behind this strategy is to train a classifier more specialised in the recovery of one specific class at the first step, and then separate the remaining classes in the second step. In the first step we fix a risk class, say \emph{L}, and we merge the other two classes into a fictitious class \emph{X}. We fit a first instance of the chosen model on the modified database. In the second step, we train another instance of the model only on the two previously merged classes. 
This is repeated for all the three risk classes. In the case of class \emph{H} being the one selected for step one, we apply SMOTE  \cite{chawla2002smote} before training, a well-known algorithm for data rebalancing.\footnote{Using \emph{SMOTE} in the 1-step classification would also be an option if the objective were to use the classifier as a first filter to detect possibly critical nodes. However, we found that the overall performance of the classifier is quite poor, especially when considering the cost of classifying as highly risky (H) a firm which is creditworthy (L).}.

\begin{figure}[htb]
\centering
\includegraphics[scale=0.4]{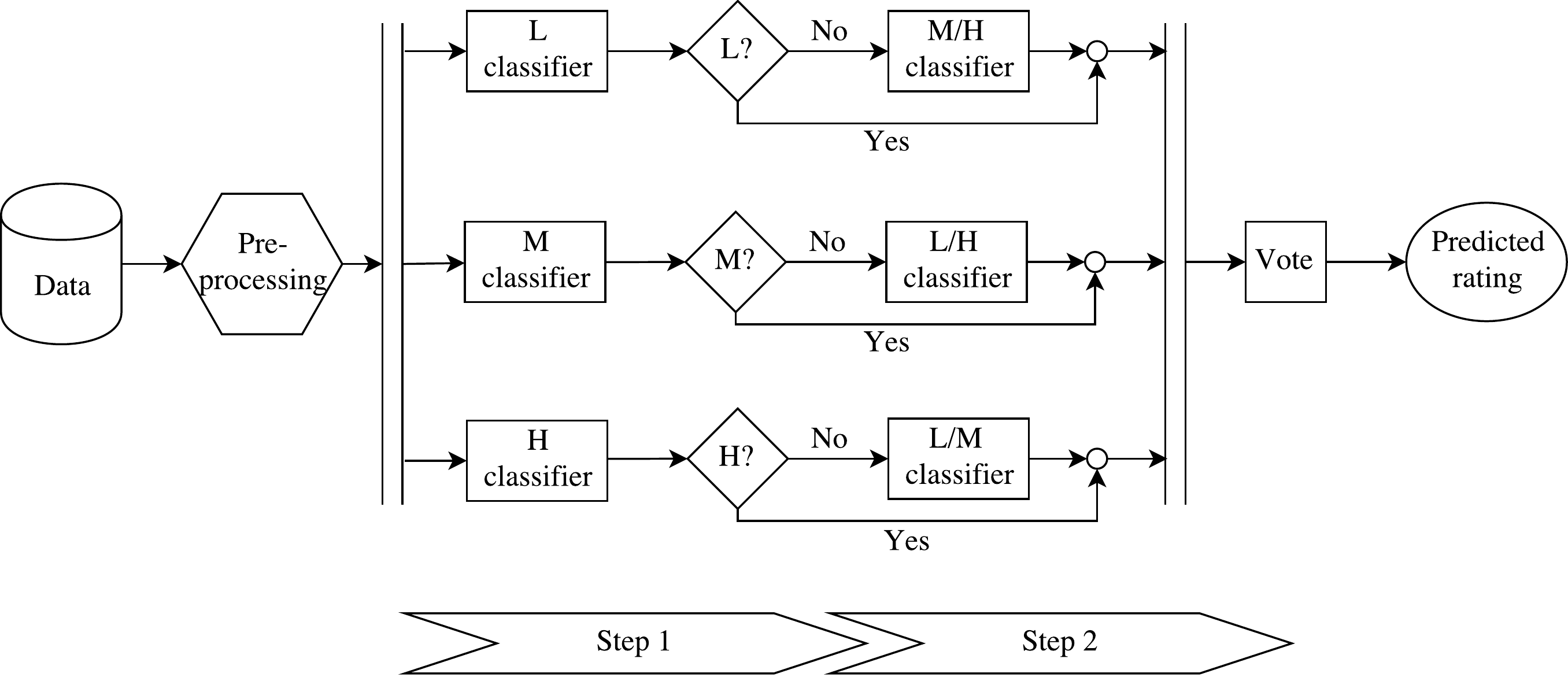}
\caption{Schematic representation of the 2-steps classifier}\label{fig:2step}
\end{figure}

Once the models are trained, the prediction are obtained by iterating the following two steps for each risk class (see the schematic representation in  Figure \ref{fig:2step})

\begin{enumerate}[noitemsep,label=\roman*.]
\item apply the first step classifier;
\item if the entry is classified as \emph{X}, apply the second step classifier.
\end{enumerate}

The final prediction is the median of the predictions. In case of draw, more weight is given when the class is obtained from the first instance (as the classifier is more specialised). For the 2-steps method, the random classifier can be defined in the following way: the null distribution for the first step is obtained for each classifier, by taking into account the  fictitious class, and at the second step by considering only the two classes previously merged.

Table \ref{tab:classification_2step} shows the results for each classifier, together with the value for the same metrics computed for the random classification. In the case of classification trees and neural networks, different combinations for the hyper-parameters have been tested (such as depth for the trees, and number and size of hidden layers for neural networks), here we present the results for the best choice for each model, and in Supplementary Information we explain the selecting procedure. 

\begin{table}[ht]
\centering
\caption{Accuracy and recall for 2-steps classifiers. R: random, ML: multinomial logistic, CT: classification tree, NN: neural network. 
}\label{tab:classification_2step}
\begin{tabular}{l|c|ccc}
method&accuracy&\multicolumn{3}{c}{recall}\\
&	&	L	&M	&	H	\\
\hline
random 1-step	&	0.413	&	0.438	&	0.625	&	0.108\\
random 2-steps 	&	0.366	&	0.368	&	0.391	&	0.249	\\
\hline
multinomial logistic	&	0.477	&	0.553	&	0.452	&	0.253	\\
classification tree&	0.496	&	0.502	&	0.567	&	0.151	\\
neural network	&	0.505	&	0.526	&	0.559	&	0.166	\\
\end{tabular}
\end{table}

The three models behave quite similarly, with slightly better overall performance of neural networks, and the training times are comparable. 
It must be noted that, among the predictors only \emph{network deduced} metrics have been included, while any data from the balance sheet, which is likely to represent the main source for the risk rating model, as well as the sector or geographic location, are excluded. When adding the economic sector, which is the only metadata available to us, as further predictor the prediction power only slightly improves to from $49\%-\,50\%$ to around $52\%$ of accuracy for both classification trees and neural networks. The natural benchmark models are the random classifiers, both 1-step and 2-steps, due to the total lack of data employed in the proprietary rating model. We are able to outperform the first by $30\%$ to $38\%$, and the latter by $15\%$ to $22\%$ in term of accuracy, and especially in the case of neural network, we are able to find a good compromise with recall for $H$.

\section{Conclusions}\label{sec:conclusions}
In this paper we empirically study the interactions and the risk distribution of 2 million Italian firms, via the investigation of payments networks built from transactional data.

Our contribution is threefold. On one side, the empirical study of the relationship between the high number of firms to our knowledge has not been done before, especially with this granularity. The study of the structure of the network highlights a complex interdependence between firms; indeed particularly interesting is the presence of a relatively small core of firms, which are involved in most transactions. This feature, paired with the power-law tail distribution of the number of connections and the total volume exchanged by the firms, can be a symptom of an architecture which favours the spread of distress, or positive feedbacks. Also relevant is the observed tendency of large, well-connected firms to be connected to small (in terms of exchanged volume), poorly connected firms. This can be the result of almost exclusive relationships between a big producer and its  subsidiaries.

The second and main contribution is the assessment of the correlation between the network structure and the distribution of risk. From our analysis, we can conclude that the risk level of a firm is related to its features and role in the network at different levels. For single firms, we observed that low risk firms are more likely to have a high number of connections, and some of them acts as hubs for the entire network, being connected to thousands of other firms. When pairs of linked firms are considered, we observed the tendency to favour connections towards firms with the same risk level. This tendency can be observed also on a more aggregate level. Indeed, we found that also groups of firms which are more connected among them that with the rest of the network, have a local distribution of risk which is statistically different from the global one, meaning that some risk classes are over- or under- represented. Finally, we divided firms into a hierarchical organisation, in such a way to highlight the main direction along which money circulates. This simplified structure showed once more that many levels of the hierarchy have a local distribution of risk statistically different from the global one. As high risk firms are over-represented at the beginning of the flow of money, this can be a source of distress for the entire system.

Finally, we showed that network metrics and communities can be successfully used to predict the missing ratings with machine learning models. We propose a simple 2-steps strategy to compromise between overall accuracy and recall on the smallest but most risky class. We test our strategy with three methods, namely multinomial logistic, classification trees and neural networks. Since predictors are all network-derived quantities, and no information from balance sheets or other meta-data are used, the random rating assignment is the natural benchmark. We find that all the three methods are able to outperform sizeably the benchmark, with slightly better results for neural networks.
\section*
%
%
%
%
%
%
%
%
%
{Acknowledgements} We would like to thank Ilaria Bordino, Francesco Gullo, Francesco Montecuccoli degli Erri, Marcello Paris and Stefano Pascolutti from R\&D team in UniCredit for useful discussions and technical support. We are also grateful for suggestions we have received from Giulia Livieri, and the participants to the Data Science Summer School at École Polytechnique in Paris, XLI AMASES Annual Meeting in Cagliari and XVIII Workshop in Quantitative Finance in Milan. 

\bibliographystyle{apalike}
\bibliography{main}

\clearpage
\appendix
\section{Dataset and network metrics}\label{a1:network}
\subsection{The dataset}\label{a:dataset}
The dataset is built from transactional data of the payment platform of a major Italian bank for a total of $47$M records.
In table \ref{tab:volume-by-status} the details of the exchanged volume by customer status is presented.
Table \ref{tab:customer_rating} shows the distribution of rating across the firms, disaggregating them in terms of their customer status.

\begin{table}[h]
\caption{Percentage of volume by customer status, the row indicates the status of the payer, the column the recipient}\label{tab:volume-by-status}
\begin{tabular}{c|cccc|c|cccc|c}
month &   no &   yes	 &  ex	&  NA  & 	    	&   no &   yes &  ex &  NA			&month \\
\hline
Jan	& 0.000 & 0.036 & 0.001 & 0.001 & 	no	& 0.000 & 0.014 & 0.000 & 0.000 	& Feb\\
	& 0.009 & 0.604 & 0.030 & 0.110 & 	yes 	& 0.027 & 0.543 & 0.037 & 0.154 \\
	& 0.000 & 0.046 & 0.002 & 0.003 & 	ex 	& 0.000 & 0.037 & 0.000 & 0.002 \\    
	& 0.000 & 0.149 & 0.002 & 0.006 & 	NA 	& 0.000 & 0.184 & 0.001 & 0.000 \\
\hline		
Mar 	& 0.000 & 0.015 & 0.000 & 0.000& 	no 	& 0.000 & 0.018 & 0.000 & 0.000 		&Apr \\           
	& 0.023 & 0.541 & 0.037 & 0.151 & 	yes 	& 0.023 & 0.525 & 0.033 & 0.155 \\          
	& 0.000 & 0.036 & 0.000 & 0.002 & 	ex 	& 0.000 & 0.040 & 0.000 & 0.002 \\           
	& 0.000 & 0.193 & 0.001 & 0.000 & 	NA 	& 0.000 & 0.199 & 0.003 & 0.000 \\
\hline	
May	& 0.000 & 0.018 & 0.000 & 0.000 & 	no 	& 0.000 & 0.015 & 0.000 & 0.000	&Jun \\           
	& 0.023 & 0.542 & 0.035 & 0.144 & 	yes	& 0.018 & 0.534 & 0.037 & 0.172 \\           
	& 0.000 & 0.040 & 0.000 & 0.001 & 	ex 	& 0.000 & 0.033 & 0.000 & 0.001 \\           
	& 0.000 & 0.194 & 0.001 & 0.000 & 	NA 	& 0.000 & 0.189 & 0.001 & 0.000 \\	
\hline
Jul 	& 0.000 & 0.014 & 0.000 & 0.000	& no 		& 0.000 & 0.014 & 0.000 & 0.000 	&Aug\\            
	& 0.019 & 0.538 & 0.031 & 0.181	& yes 	& 0.018 & 0.591 & 0.029 & 0.140 \\                     
	& 0.000 & 0.031 & 0.000 & 0.002 	& ex 		& 0.000 & 0.029 & 0.000 & 0.001 \\           
	& 0.000 & 0.183 & 0.001 & 0.000 	& NA 	& 0.000 & 0.172 & 0.005 & 0.000 \\
\hline	
Sep	 & 0.000 & 0.015 & 0.000 & 0.000	& no & 0.000 & 0.013 & 0.000 & 0.000 &Oct\\           
	 & 0.019 & 0.599 & 0.027 & 0.131 	& yes & 0.022 & 0.581 & 0.029 & 0.141 \\                    
	 & 0.000 & 0.032 & 0.000 & 0.001 	& ex & 0.000 & 0.037 & 0.000 & 0.001 \\                    
	 & 0.000 & 0.175 & 0.001 & 0.000 	& NA & 0.000 & 0.175 & 0.000 & 0.000 \\	
\hline	
Nov	 & 0.000 & 0.015 & 0.000 & 0.000	& no & 0.000 & 0.014 & 0.000 & 0.000 &Dec\\                     
	& 0.013 & 0.578 & 0.037 & 0.165	& yes & 0.012 & 0.578 & 0.036 & 0.194 \\                    
	& 0.000 & 0.031 & 0.000 & 0.001 	& ex & 0.000 & 0.028 & 0.001 & 0.001 \\                   
	& 0.000 & 0.158 & 0.000 & 0.000 	& NA & 0.000 & 0.137 & 0.000 & 0.000 
\end{tabular}
\end{table}

\begin{table}[h]
\centering
\caption{Average monthly distribution of nodes by customer status and rating.}
\begin{tabular}{c|c|ccc}
status&rating&count&$\%$&$\%$with rating	\\
\hline
 			&L&2121&0.000&\\
not customer	&M&	4592&0.003&0.010		\\
incl. NA	&H&	305	&0.000&		\\
			&ND&676762&0.990	&		\\
\hline
\multirow{ 4}{*}{customer}		&L&87801&0.305&\\
			&M&95893&	0.333	&0.702	\\
			&H&	18811&	0.065	&		\\
			&ND&85841&	0.298	&		\\
\hline
\multirow{ 4}{*}{former}
&L&3901	&	0.017	&		\\
			&M&7850	&	0.179	&0.340		\\
			&H&	926	&	0.017	&		\\
			&ND&41775	&	0.767	&		\\
\hline
total		&&	1026577	&&	0.217	\\
\end{tabular}\label{tab:customer_rating}
\end{table}


\subsection{Time aggregation}\label{a:time}
When defining a network from temporal data, choosing the time scale of analysis is crucial because it can affect deeply the topology. 
Shorter time scales (daily or weekly) emphasise peculiar behaviours as, for example, which supplier is paid first once liquidity is available. Longer time scales help giving a more stable picture of the supply chain structure among firms.

\begin{figure}[htb]
\setlength\figureheight{0.2\columnwidth} 
\setlength\figurewidth{0.2\columnwidth} 
%
%
\definecolor{nodes}{RGB}{155, 174, 209}%
\definecolor{edges}{RGB}{159, 201, 173}%

\begin{tikzpicture}

\begin{axis}[%
width=\figurewidth,
height=\figureheight,
scale only axis,
axis on top,
xmin=0,
xmax=365,
xlabel={days},
ymin=1,
ymax=100000000,
ymode = log,
ylabel={frequency},
name=day,
]
\addplot [forget plot] graphics [xmin=0,xmax=400,ymin=1,ymax=100000000] {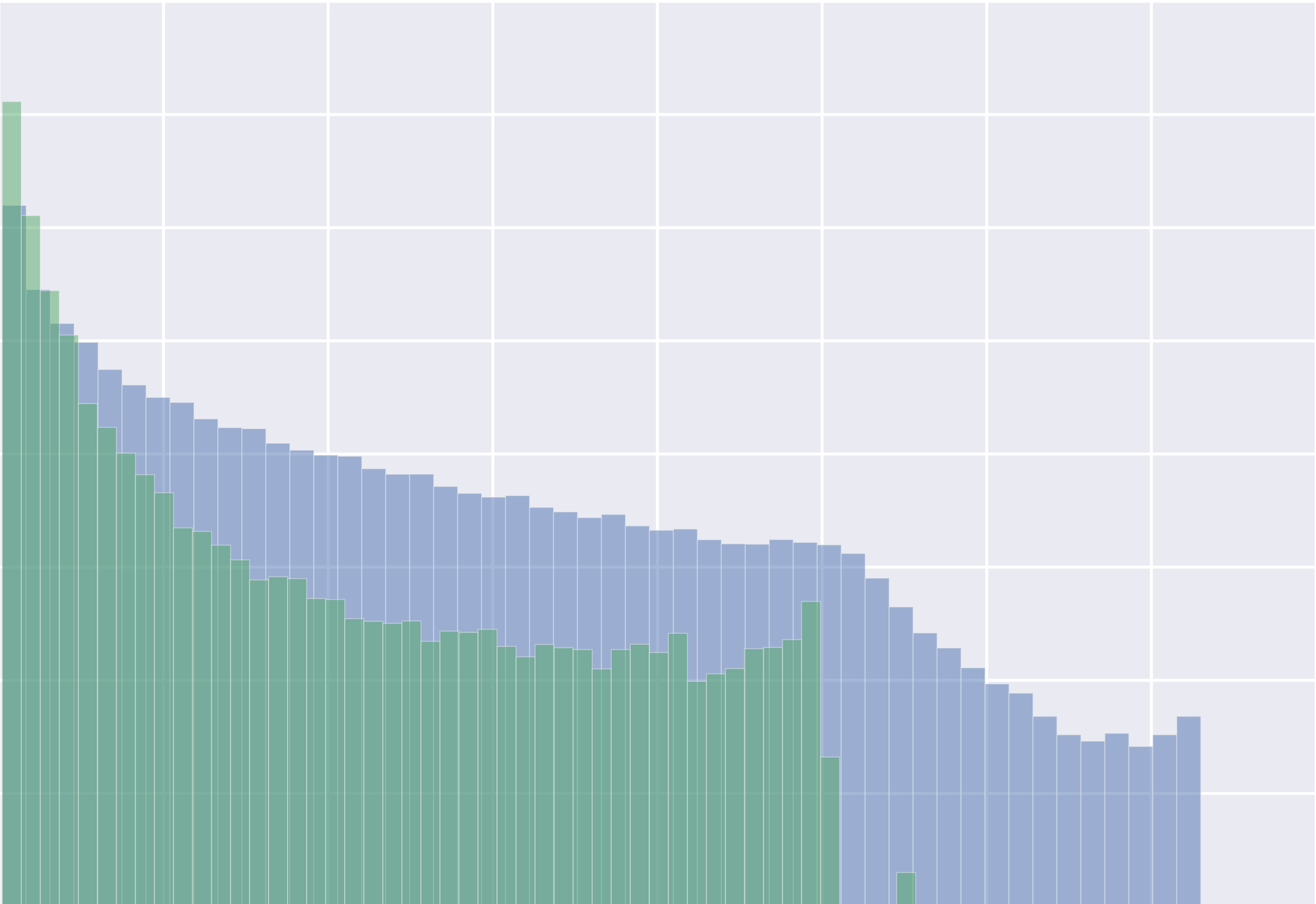};
\end{axis}

\begin{axis}[%
width=\figurewidth,
height=\figureheight,
scale only axis,
axis on top,
xmin=0.5,
xmax=52.5,
xtick={10,20,30,40,50},
xlabel={weeks},
ymin=100,
ymax=10000000,
ymode = log,
name=week,
at =(day.right of south east),
anchor=left of south west,
]
\addplot [forget plot] graphics [xmin=0,xmax=60,ymin=100,ymax=10000000] {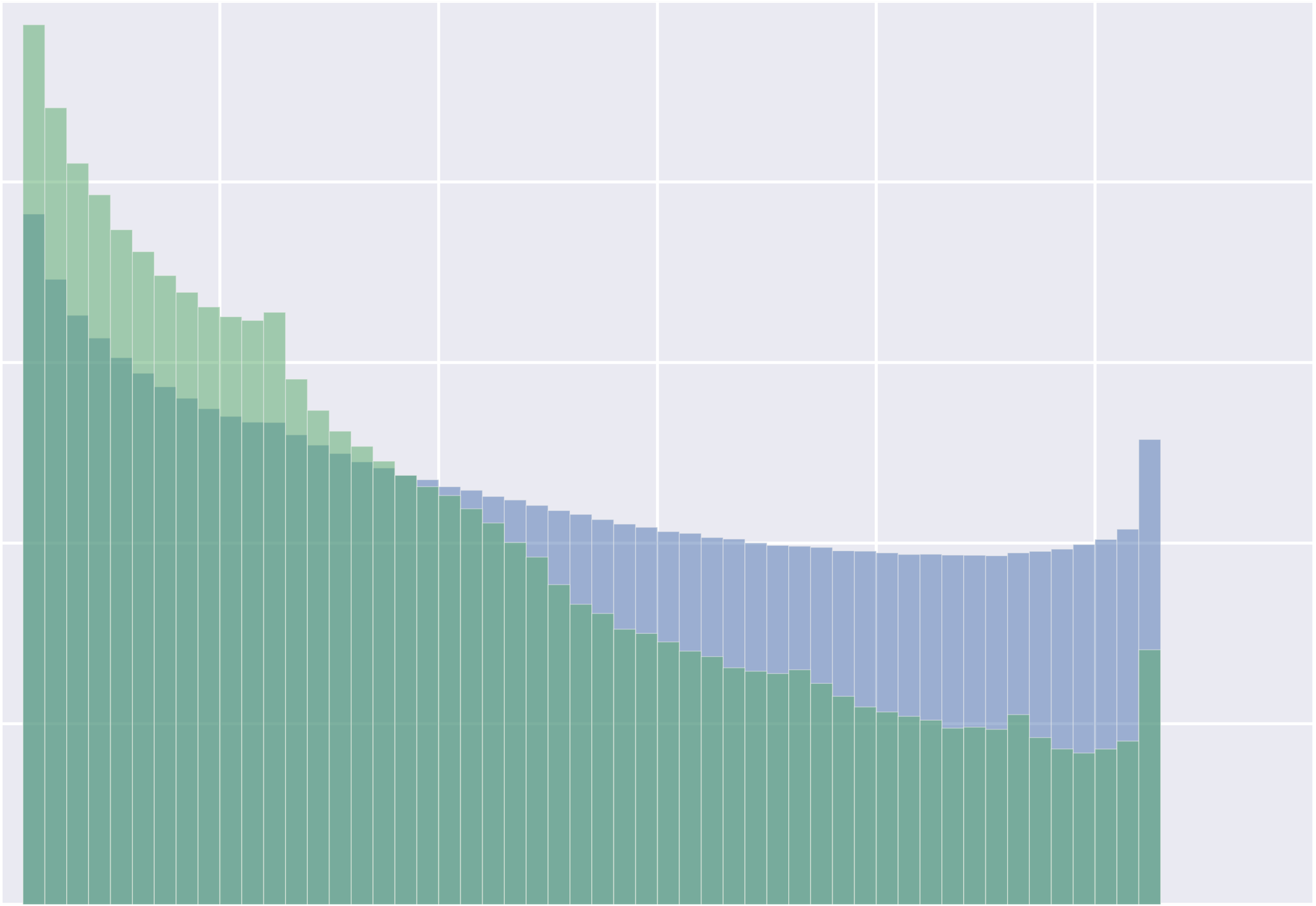};
\end{axis}

\begin{axis}[%
width=\figurewidth,
height=\figureheight,
scale only axis,
axis on top,
xmin=0.5,
xmax=12.5,
xtick={1,6,12},
xlabel={months},
ymin=10000,
ymax=10000000,
ymode = log,
legend entries={nodes,links},
legend style={area legend,legend cell align=left,align=left,draw=none},
legend pos = {outer north east},
name=month,
at =(week.right of south east),
anchor=left of south west,
]
\addplot [forget plot] graphics [xmin=-0.5,xmax=12.5,ymin=10000,ymax=10000000] {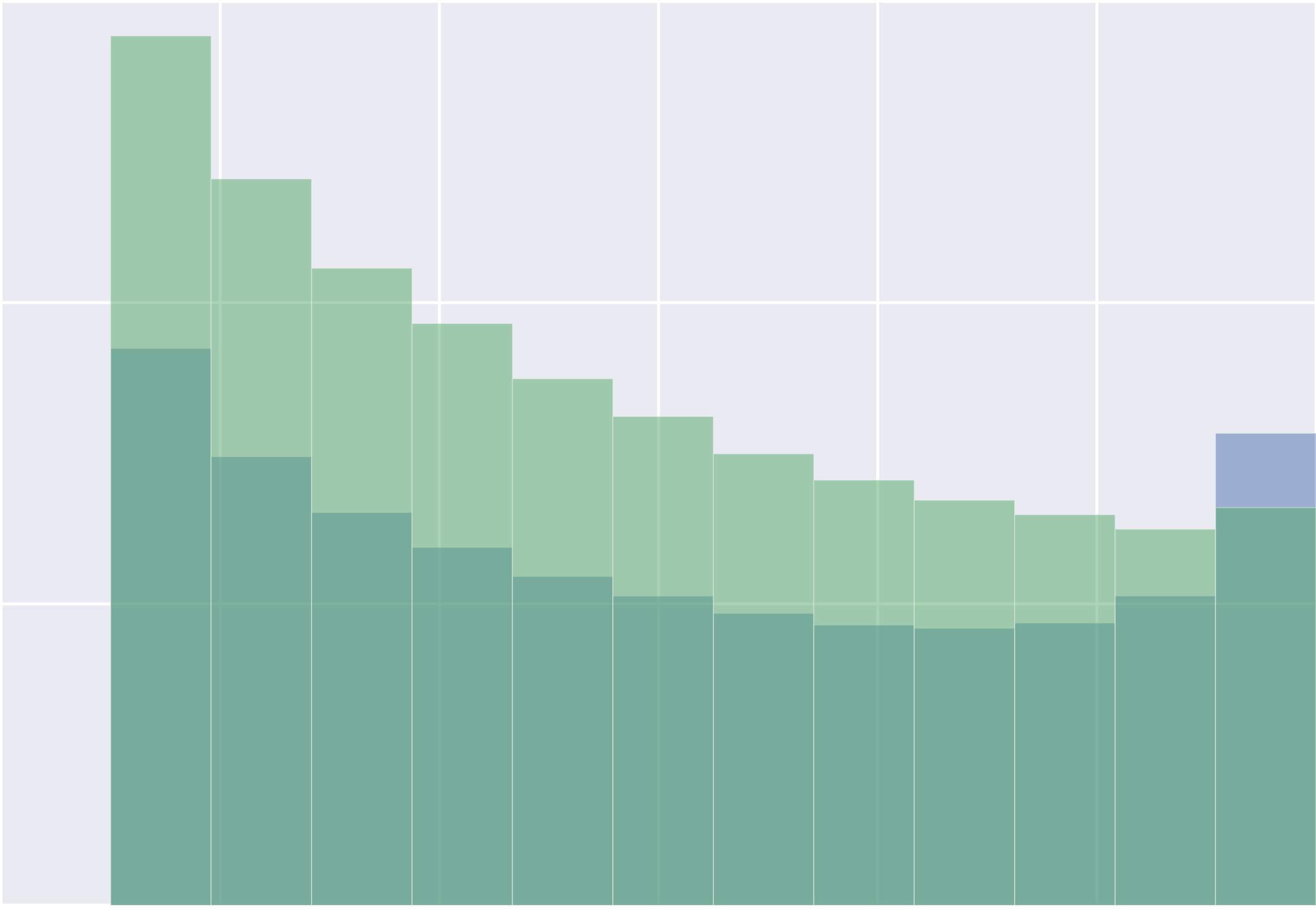};
\addlegendimage{no markers,fill = nodes}
\addlegendimage{no markers,fill = edges}
\end{axis}

\end{tikzpicture}%
\caption{Histogram for the number of days, weeks, months of activity for nodes (blue) and existence for edges (green) for different time aggregations}\label{fig:pers}
\end{figure}

\begin{figure}[p]
\setlength\figureheight{0.3\columnwidth} 
\setlength\figurewidth{0.7\columnwidth} 
\centering
\begin{tikzpicture}

\begin{axis}[%
width=\figurewidth,
height=\figureheight,
scale only axis,
axis on top,
xmin=1,
xmax=365,
xtick={1,32,60,91,121,152,182,213,244,274,305,335},
xticklabels={Jan,Feb,Mar,Apr,May,Jun,Jul,Aug,Sep,Oct,Nov,Dec},
xlabel={days},
ymin=1,
ymax=1200000,
name=day,
]
\addplot [forget plot] graphics [xmin=1,xmax=365,ymin=1,ymax=1200000] {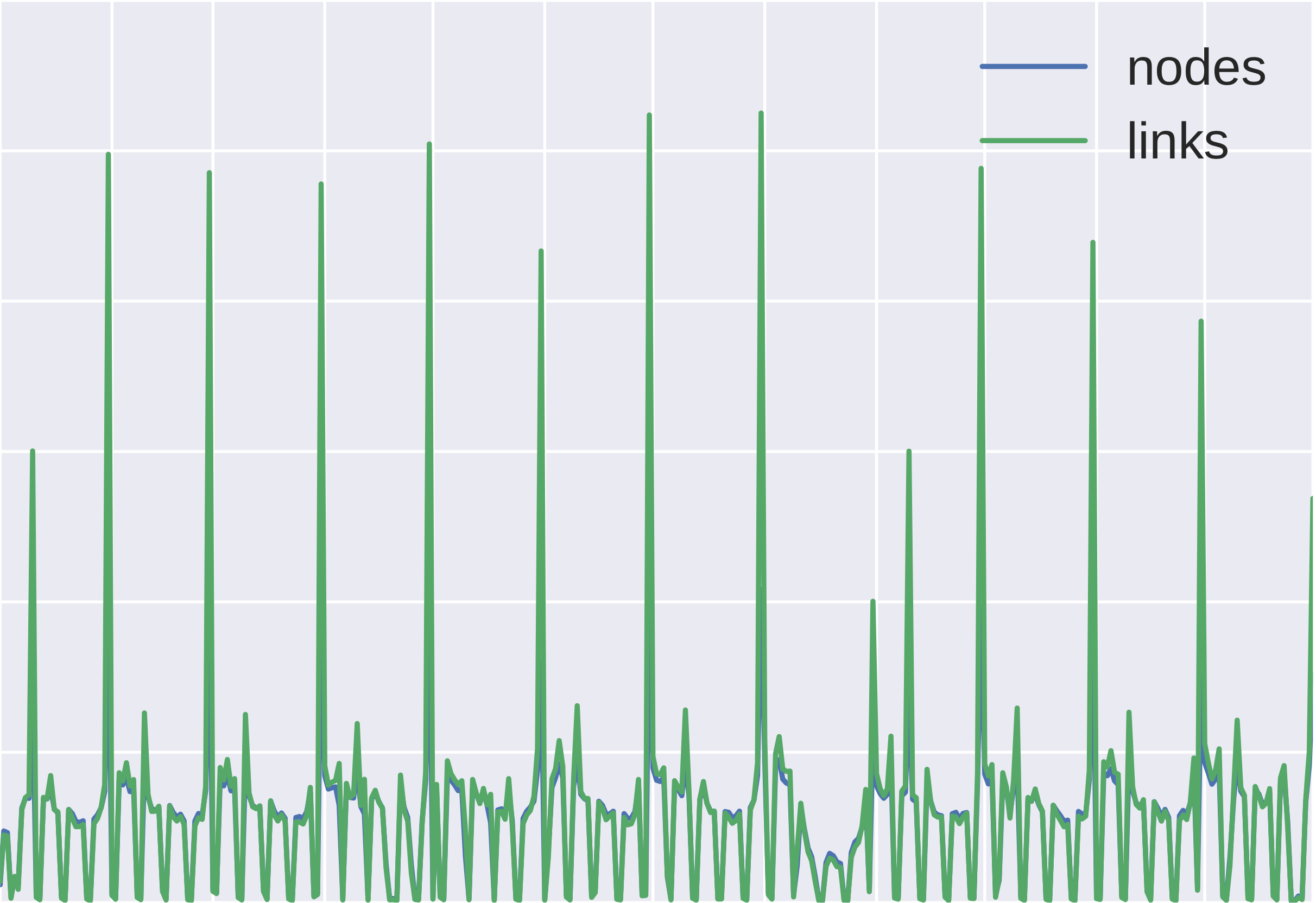};
\end{axis}

\begin{axis}[%
width=\figurewidth,
height=\figureheight,
scale only axis,
axis on top,
xmin=1,
xmax=365,
xtick={1,32,60,91,121,152,182,213,244,274,305,335},
xticklabels={Jan,Feb,Mar,Apr,May,Jun,Jul,Aug,Sep,Oct,Nov,Dec},
xlabel={weeks},
ymin=200000,
ymax=1800000,
name=week,
at =(day.below south east),
anchor=above north east,
]
\addplot [forget plot] graphics [xmin=1,xmax=365,ymin=200000,ymax=1800000] {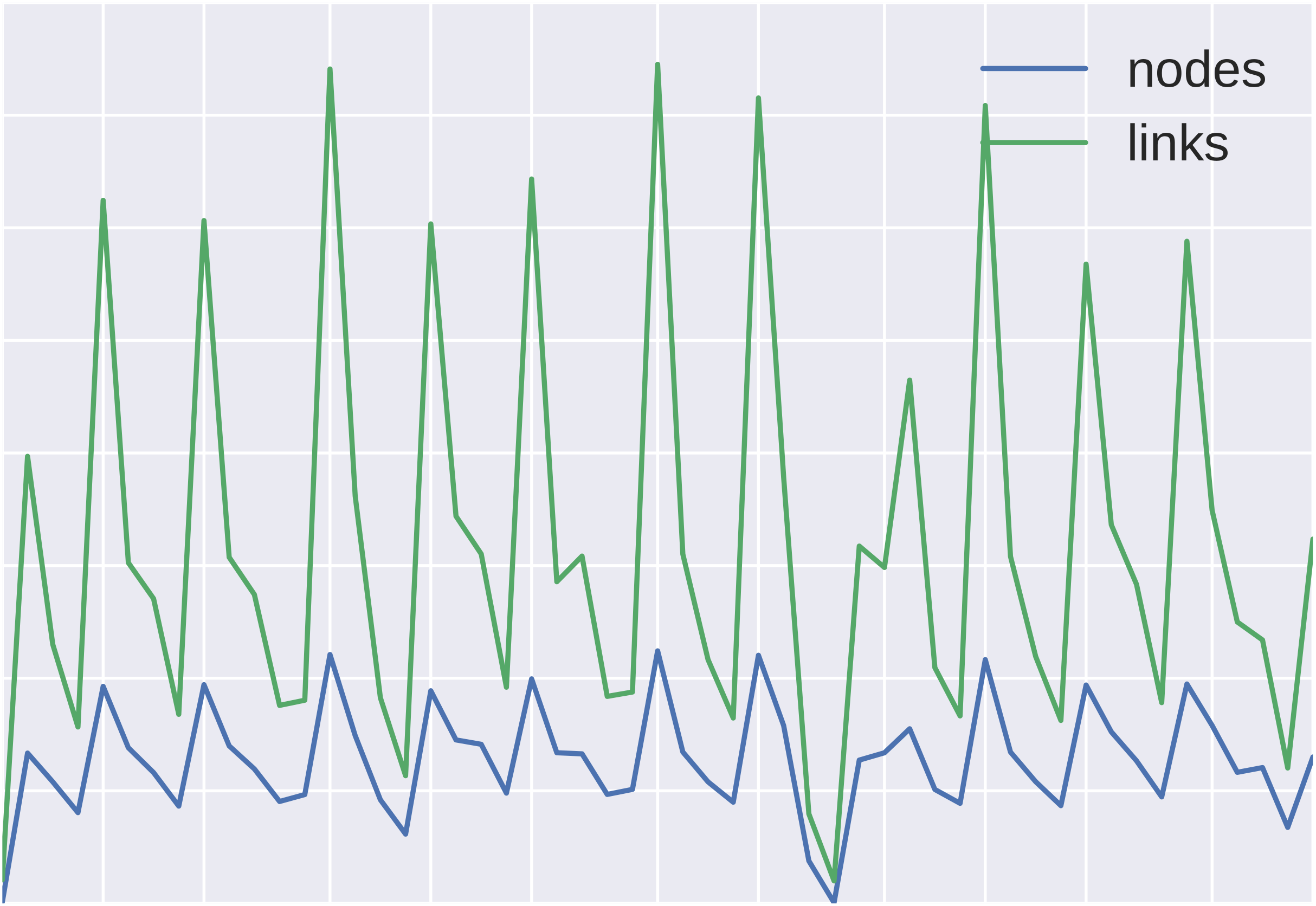};
\end{axis}

\begin{axis}[%
width=\figurewidth,
height=\figureheight,
scale only axis,
axis on top,
xmin=1,
xmax=365,
xtick={1,32,60,91,121,152,182,213,244,274,305,335},
xticklabels={Jan,Feb,Mar,Apr,May,Jun,Jul,Aug,Sep,Oct,Nov,Dec},
xlabel={months},
ymin=500000,
ymax=4000000,
name=month,
at =(week.below south east),
anchor=above north east,
]
\addplot [forget plot] graphics [xmin=1,xmax=365,ymin=500000,ymax=4000000] {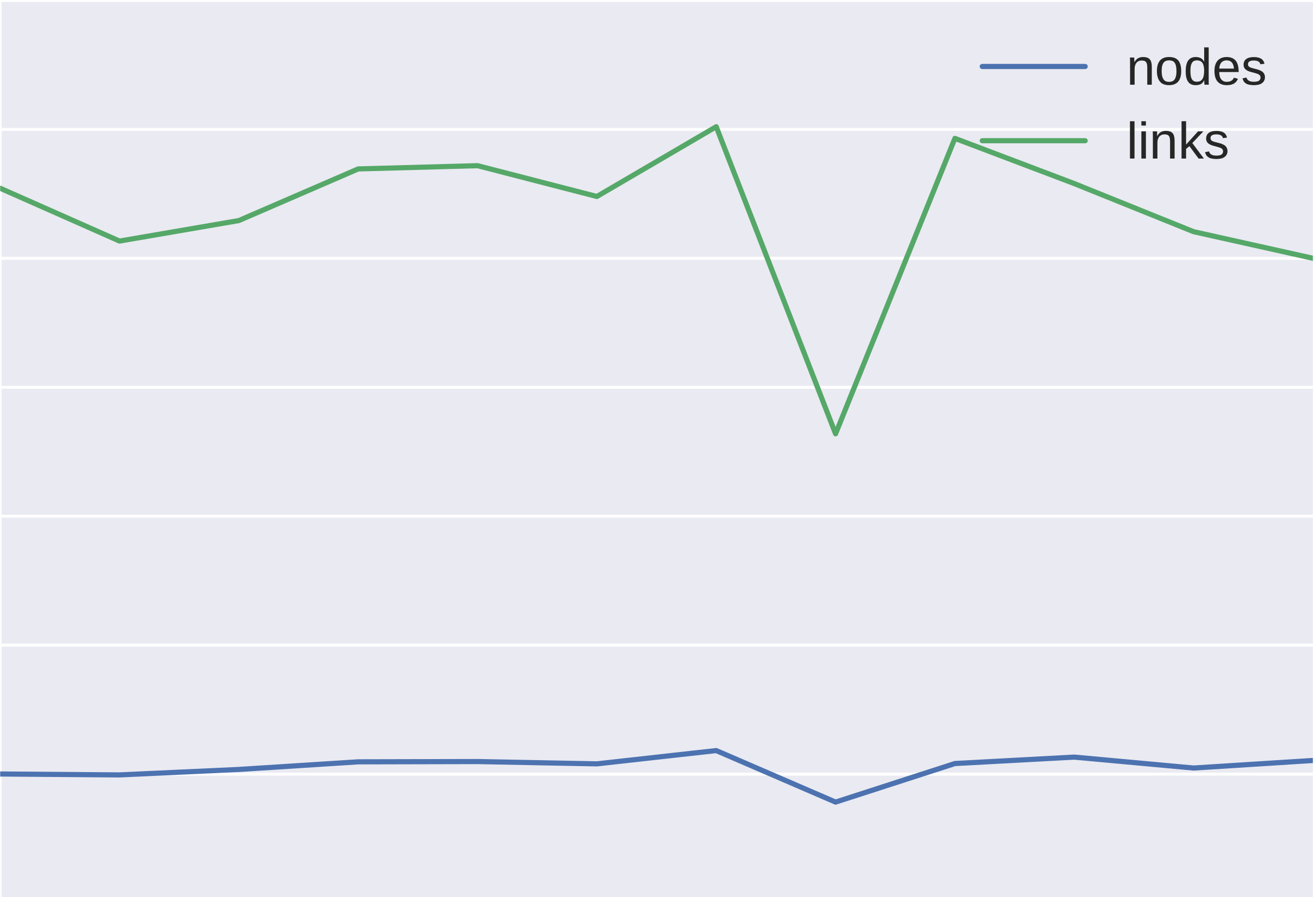};

\end{axis}

\end{tikzpicture}%
\caption{Number of nodes (blue) and links (green) for each day, week, and month. Only with longer time aggregations one is able to eliminate periodicity.}\label{fig:size}
\end{figure}

In order to give an intuition of different behaviours, two quantities can be considered. The first is the persistence of links and nodes, which is measured by counting the number of times a node or an edge appears in the networks for different time aggregations. From Figure \ref{fig:pers} one can see that most of nodes are active only for few days, while a small core of firms is intensely active through the whole year. 
Secondly, the size of the networks, both in terms of number of nodes and links, for different time aggregations is shown in Figure \ref{fig:size}. Interestingly, for daily aggregation, see Left panel, both quantities show a high periodicity, with a very high peak (a factor $\sim5$  with respect to the other days) at the end of each month. This effect is evident also with weekly aggregation, see (central panel), but not in the monthly time scale. This last observation justifies the choice of monthly networks as focus of this analysis.

\clearpage
\subsection{Network metrics}\label{a:further}
A network's component is a subsets of nodes such that there is path between any pair of nodes, either undirected (weakly connected components), or directed (strongly connected components). From the definition of the networks it is clear that there are no isolated nodes, since the smallest weak components include at least two nodes, namely a payer and a payee. As it is common for many other real networks, it is possible to identify a weak component, which is of the order of magnitude of the entire network. In our case this giant component (GC) includes on average $98\%$ of the nodes. Considering instead the largest strongly connected component (SCC), it includes approximately $20\%$ of the nodes but more than  half of the links. As a consequence the density of the strongly connected component is an order of magnitude larger than the density of the whole network or of the weakly connected component. See Table \ref{tab:comp} in for more details of these quantities. 

In the standard definition of the bow-tie structure of a network, the nodes in the GC but outside the strongly connected component are divided between the in-component, the nodes from which links arrive in the strongly connected component, and the out-component, the nodes reachable from the SCC. 
Nodes in the in-component that have no incoming links, represent each month about one half of the active firms and their activity is sporadic. 

\begin{table}[h]
\centering
\caption{Percentage size ($\%n$) and density ($\rho$) of the largest weakly (GC) and strongly (SCC) connected components. The last column ($\%w$) contains the relative volume transferred among nodes in the SCC with respect to the total volume.}\label{tab:comp}
\begin{tabular}{l|cc|ccc}
&\multicolumn{2}{c|}{  GC}&\multicolumn{3}{c}{SCC}\\
 month &          $\%n$   &     $\rho$ & $\%n$   &     $\rho=\frac{m}{n(n-1)}$& $\%w$  \\
 \hline
Jan &    0.989 &  $3.4\cdot10^{-6}$  &  0.232 & $3.29  \cdot10^{-5}$&  0.75 \\
Feb &    0.989 &  $3.21\cdot 10^{-6}$&0.237   &  $2.99\cdot 10^{-5}$ &  0.69\\
Mar &    0.980 & $  3.15\cdot 10^{-6 }$ &0.235 &  $2.98\cdot 10^{-5 }$&  0.70 \\
Apr &    0.980 & $ 3.16\cdot 10^{-6 }$&  0.231 &  $3.09\cdot 10^{-5}$&  0.67 \\
May &    0.981 &$   3.16\cdot 10^{-6 }$&  0.232 &$ 3.06\cdot 10^{-5}$ &  0.69\\
Jun &    0.980  &  $3.11\cdot 10^{-6 }$&  0.230&  $3.03\cdot 10^{-5}$&  0.69 \\
Jul &   0.982 &  $ 3.05\cdot 10^{-6 }$& 0.237 & $2.88\cdot 10^{-5}$&  0.70 \\
Aug &    0.970 &$  3.08\cdot 10^{-6}$ &  0.204 & $ 3.23\cdot 10^{-5}$ &  0.69\\
Sep &   0.981 &$ 3.31\cdot 10^{-6 }$&  0.233 &  $3.23\cdot 10^{-5}$ &  0.73\\
Oct &   0.981 & $ 3.00\cdot 10^{-6  }$&  0.237 &  $2.81\cdot 10^{-5}$&  0.68 \\
Nov &    0.979 &$ 3.08\cdot 10^{-6 }$&  0.227 & $3.00\cdot 10^{-5}$ &  0.65 \\
Dec &   0.979 &$  2.81\cdot 10^{-6  }$&  0.228 & $2.69\cdot 10^{-5}$&  0.67 
  \end{tabular}

\end{table}

\begin{table}\centering
\caption{Results of power law fit of the degree and strength distribution for all the months obtained by using the algorithm described in \cite{clauset2009power}. The $\alpha$ parameter is the fitted exponent and the $k_{min}$ and $w_{min}$ parameter is the estimated minimum value after which the behaviour of the distribution is consistent with a power law tail. Since the volume of payments are scaled, the values of $w_{min}$s are not much informative, so for the strength $F(w_{min})=1-EDCF(w_{min})$ is reported instead. } \label{tab:pl-deg}
\begin{tabular}{c|cccc|cccc}

&\multicolumn{4}{c|}{degree}&\multicolumn{4}{c}{strength}\\
month&$\alpha^{\text{in}}$&$k_{\text{min}}^{\text{in}}$&$\alpha^{\text{out}}$&$k_{\text{min}}^{\text{out}}$
&$\alpha^{\text{in}}$&$F(w_{\text{min}}^{\text{in}})$&$\alpha^{\text{out}}$&$F(w_{\text{min}}^{\text{out}})$\\
\hline
Jan&2.55&159&2.85&24&1.89&0.03&1.89&0.01\\
Feb&2.56&148&2.80&19&2.10&0.03&1.99&0.01\\
Mar&2.53&125&2.70&44&2.11&0.03&1.97&0.01\\
Apr&2.67&257&2.82&22&2.14&0.03&2.01&0.01\\
May&2.66&227&2.83&23&2.12&0.03&2.07&0.01\\
Jun&2.58&135&2.83&23&2.07&0.03&1.96&0.01\\
Jul&2.52&124&2.80&21&2.07&0.03&	2.02&	0.01\\
Aug&2.62&236&2.74&14&2.07&0.03&	1.94	&0.01\\
Sep&2.64&187&2.83&32&2.09&	0.04&2.03	&0.01\\
Oct&2.53&129&2.75&25&2.05&	0.03	&1.95	&0.01\\
Nov&2.59&134&2.83&19&2.06&0.03	&2.04	&0.01\\
Dec&2.55&180&2.62&41&2.06	&0.03	&1.98	&0.01\\
\end{tabular}

\end{table}

\begin{table}
\centering

\caption{Assortativity coefficient for degree and strength. The columns \emph{with rating} refers to the subgraph of nodes with known rating. The columns \emph{customers} refers to the subgraph of nodes with customer status \emph{yes}. }\label{tab:assort}
\begin{tabular}{c|ccc|ccc}
attribute&\multicolumn{3}{c|}{degree}&\multicolumn{3}{c}{strength}\\
\hline
nodes	&	all	&	with rating	&	clients	&	all	&	with rating	&	clients\\
\hline
Jan	&	-0.035	&	-0.035	&	-0.046	&	-0.036	&	-0.031	&	-0.046	\\
Feb	&	-0.027	&	-0.029	&	-0.036	&	-0.030	&	-0.031	&	-0.039	\\
Mar	&	-0.025	&	-0.030	&	-0.038	&	-0.027	&	-0.027	&	-0.037	\\
Apr	&	-0.027	&	-0.027	&	-0.037	&	-0.030	&	-0.029	&	-0.041	\\
May	&	-0.026	&	-0.033	&	-0.039	&	-0.027	&	-0.026	&	-0.035	\\
Jun	&	-0.025	&	-0.027	&	-0.032	&	-0.028	&	-0.030	&	-0.037	\\
Jul	&	-0.025	&	-0.028	&	-0.036	&	-0.028	&	-0.028	&	-0.038	\\
Aug	&	-0.027	&	-0.035	&	-0.040	&	-0.032	&	-0.034	&	-0.041	\\
Sep	&	-0.024	&	-0.027	&	-0.030	&	-0.028	&	-0.030	&	-0.036	\\
Oct	&	-0.028	&	-0.028	&	-0.037	&	-0.032	&	-0.033	&	-0.043	\\
Nov	&	-0.023	&	-0.028	&	-0.031	&	-0.026	&	-0.028	&	-0.035	\\
Dec	&	-0.027	&	-0.030	&	-0.034	&	-0.031	&	-0.035	&	-0.041	\\
\end{tabular}

\end{table}
\clearpage

\section{Risk distribution}
\subsection{Degree and risk}\label{a:deg_risk}
The multinomial logistic regression aims to model the probabilities for a classification problem with more than two outcomes. Here we treat the responses ($L,\,M,\,H$) as categorical and ordered.
In practice this means to find parameters that best fit the model
\[\log\left(\frac{P(r\leq L)}{P(r> L)}\right)=a_L + b_L^1 X_1 ...+b_L^p X_p+ 
\]
\[\log\left(\frac{P(r\leq M)}{P(r> M)}\right)=a_L + b_M^1 X_1 ...+b_M^p X_p+ 
\].
$X_i$ are the predictors, $a$ and $b_\cdot^i$ are the coefficients.
We consider the cases $p=1$, where the predictor is the degree $X_1=k$, and the case $p=2$ where also the size is used as predictor $X_2=s$. In the following table \ref{tab:logit_results}, the $b$ coefficients are shown, together with an indication for the statistical significance.

\begin{table}[htb]
\centering
\caption{Coefficients for multinomial logistic regression. The first two columns refer to the regression with the degree as only predictor. 
The last four columns refer to the regression with also the size as predictor. The 
superscript indicates the predictors: $k$ for the degree, $s$ for the size. The subscript indicates the risk rating. The stars indicate significance: one star if the p-value $<0.05$, two stars if the p-value $<0.01$}\label{tab:logit_results}
\begin{tabular}{c|ll|llll}
	&	$b_L^k$		&	$b_M^k$		&	$b_L^k$		&	$b_L^s$		&	$b_M^k$		&	$b_M^s$		\\
    \hline
Jan	&	0.258	**	&	0.352	**	&	0.261	**	&	-0.007		&	0.312	**	&	0.091	**	\\
Feb	&	0.221	**	&	0.305	**	&	0.210	**	&	0.024	**	&	0.233	**	&	0.159	**	\\
Mar	&	0.226	**	&	0.314	**	&	0.220	**	&	0.013		&	0.237	**	&	0.173	**	\\
Apr	&	0.243	**	&	0.328	**	&	0.240	**	&	0.007		&	0.270	**	&	0.131	**	\\
May	&	0.229	**	&	0.324	**	&	0.206	**	&	0.051	**	&	0.237	**	&	0.195	**	\\
Jun	&	0.239	**	&	0.325	**	&	0.232	**	&	0.017	*	&	0.237	**	&	0.199	**	\\
Jul	&	0.238	**	&	0.344	**	&	0.227	**	&	0.026	**	&	0.263	**	&	0.187	**	\\
Aug	&	0.183	**	&	0.272	**	&	0.175	**	&	0.020	*	&	0.179	**	&	0.211	**	\\
Sep	&	0.238	**	&	0.355	**	&	0.218	**	&	0.046		&	0.255	**	&	0.228	**	\\
Oct	&	0.220	**	&	0.329	**	&	0.207	**	&	0.030	**	&	0.232	**	&	0.220	**	\\
Nov	&	0.226	**	&	0.338	**	&	0.211	**	&	0.034	**	&	0.233	**	&	0.234	**	\\
Dec	&	0.219	**	&	0.331	**	&	0.220	**	&	-0.002	*	&	0.231	**	&	0.227	**	\\

\end{tabular}
\end{table}
\clearpage
\subsection{Assortativity of risk}\label{a:assort}

\begin{table}[h]
\centering
\caption{Assortativity coefficient for risk rating. The columns \emph{with rating} refers to the subgraph of nodes with known rating. The columns \emph{customers} refers to the subgraph of nodes with customer status \emph{yes}. In the last two columns, the metric for assortativity is modified in order to take into account weights, specifically  $e_{ij}$ is computed as the fraction of volume, not the number of edges (see main text for more details). }\label{tab:assort_risk}
\begin{tabular}{c|ccc|ccc}
metric& \multicolumn{3}{c|}{standard}&\multicolumn{3}{c}{weighted}\\
\hline
nodes	&	all	&	with rating	&	clients	&	all	&	with rating	&	clients\\
\hline
Jan	&	-0.063	&	0.025	&	0.035	&	0.073	&	0.115	&	0.109\\
Feb	&	-0.066	&	0.026	&	0.038	&	0.106	&	0.181	&	0.188\\
Mar	&	-0.067	&	0.025	&	0.039	&	0.073	&	0.150	&	0.150\\
Apr	&	-0.067	&	0.026	&	0.036	&	0.069	&	0.154	&	0.156\\
May	&	-0.067	&	0.025	&	0.038	&	0.065	&	0.146	&	0.139\\
Jun	&	-0.068	&	0.026	&	0.039	&	0.060	&	0.150	&	0.128\\
Jul	&	-0.072	&	0.025	&	0.037	&	0.046	&	0.142	&	0.137\\
Aug	&	-0.078	&	0.025	&	0.040	&	0.078	&	0.149	&	0.224\\
Sep	&	-0.067	&	0.025	&	0.040	&	0.087	&	0.168	&	0.216\\
Oct	&	-0.076	&	0.024	&	0.037	&	0.080	&	0.151	&	0.213\\
Nov	&	-0.072	&	0.024	&	0.039	&	0.070	&	0.175	&	0.149\\
Dec	&	-0.082	&	0.024	&	0.040	&	0.037	&	0.199	&	0.151
\end{tabular}
\end{table}

To test if nodes show different preferences in connection between incoming and outgoing payments we define the quantities
\begin{align*}
\Delta_i^{(\text{in})}(X)&=\frac{w_i^{(\text{in})}(X)-\tilde{a}_X\tilde{b}_{r(i)}}{1-\tilde{a}_X\tilde{b}_{r(i)}}\,,\quad X\in\{L,\,M,\,H\}\\
\Delta_i^{(\text{out})}(X)&=\frac{w_i^{(\text{out})}(X)-\tilde{a}_{r(i)}\tilde{b}_X}{1-\tilde{a}_{r(i)}\tilde{b}_X}\,,\quad X\in\{L,\,M,\,H\}\,.
\end{align*}
The notation is consistent with the  definition in \eqref{eq:assort_coeff}: $r(i)$ is the risk of node $i$; $\tilde{a}_X,\tilde{b}_X$ are the percentage volume from or to nodes with rating $X$ for the whole network, $w_i^{(\text{out})}(X)$ ($w_i^{(\text{in})}(X)$)  is the percentage of the volume from (to) node $i$ to (from) nodes of rating $X$. Samples are obtained by grouping nodes by rating, for a total of $18(=(3\text{ ratings})^2\cdot2\text{ directions})$ distributions. For example, the distribution of excess percentage volume from $L$ towards $M$ is given by 
\[
\{\Delta_i(M)^{(\text{out})}\,|\,i\in L\}\sim F^{(\text{out})}_L(M)\,.
\]
Similarly, the excess percentage volume entering $M$ from $H$ is given by 
\[
\{\Delta_i(H)^{(\text{in})}\,|\,i\in M \}\sim F^{(\text{in})}_M(H)\,.
\]
Note that in general, $F^{(\text{in})}_X(Y)\neq F^{(\text{in})}_Y(X)$.

We perform two set of test. In the first case we fix one rating and we compare out- and in- excess percentage volume with respect to a certain rating. In all the cases the null hypothesis is rejected with very low p-values, however it is not straightforward to give an economic interpretation of the overall results: for all the rating, the excess percentage toward $L$ is greater that the analogous for incoming volume, while the opposite holds for payments to and from $H$.
In the second set of test we fix a rating and a direction (in or out), and we compare the excess percentage volume from (or to) all the ratings. Also in this case all the tests reject the null with very low p-values, so we are able to order the distributions and evaluate the preference in connection. For the outgoing volume, rating $L$ is preferred to the more risky ones in all the case. Payments to nodes rated $M$ follows in preference from nodes having risk $M$ and $H$, but are last in order for nodes having rating $L$. For incoming payments, the situation is slightly different. Rating $M$ is preferred by nodes rated $M$ and $H$, and it is followed by $L$. While the preference is reversed for payments from nodes rated $L$. 

\subsection{Test for risk distribution within a community}\label{a:test}



The statistical test employed in the main text has the purpose to assess whether a given rating is under- or over- represented in a certain subset, obtained by one of the partitioning methods described in the paper.  In general, this means to test if the distribution of ratings in a single subset is statistically different from the unconditional distribution obtained considering the entire sample. To do so, one computes the p-value representing the probability to observe a given number of ratings in each community under the null hypothesis of that ratings are distributed in the community as in the whole sample. 
As shown in \cite{tumminello2011community} the probability under the null is the hyper-geometric distribution. Moreover, since for each community multiple tests (one for each rating and community) are performed, a correction for the p-value for multiple hypothesis testing is used. In particular, the Bonferroni correction is chosen, i.e. fixed a threshold $p_s$ for the p-value, the corrected threshold is given by $\frac{p_s}{N_r}$, where $N_r$ is the number of tests. The threshold of is fixes at $p_s=1\%$ before correction.

Specifically, given a partition $\{C_i\}_i$ the following quantities are computed
\begin{align*}
k_{x,i}&=\#\{\text{nodes in $C_i$ with rating $x$}\}\\
n_i&=\#\{\text{nodes in $C_i$}\}\\
K_x&=\#\{\text{nodes with rating $x$}\}\\
N'&=\#\{\text{nodes}\}
\end{align*}
and the p-value is given by
\[
p=\begin{cases}\mathbb{P}(y>k_{x,i}&\frac{k_{x,i}}{n_i}>\frac{K_x}{N'})\\
\mathbb{P}(y<k_{x,i}&\frac{k_{x,i}}{n_i}<\frac{K_x}{N'})
\end{cases}\,,\quad y\sim \text{hypergeom}\left(\frac{K_L}{N'},\frac{K_M}{N'},\frac{K_H}{N'}; N'\right)\,.
\]

Note that $\{K_x\}$ and $N'$ are computed in the specific monthly network under consideration.

In the case of the distribution conditioned on the distance, the subsets are obtained by considering pairs of nodes.
For example, the fraction of nodes with rating L at distance $k$ from H is computed as
\[
p^{(k)}_{HL}=\frac{\abs{\{(i,j): d(i,j)=k, \,i\in\, H,\,j\in\,L\}}}{\abs{\{(i,j): d(i,j)=k,i\in\, H\}}}\,.
\]

The partitions resulting from the other methods are very different in terms of number and size of subsets, so to make tests comparable, only communities including at least 500 nodes with known rating. In the cases of modularity, subsets are ordered by descending size. Note that, since each month the set of active nodes and the labelling of subsets changes, one cannot easily compare the behaviour of a subsets across months. 

Tables \ref{tab:test_mod_sum}, \ref{tab:test_ag_sum}, present a summary of the tests, recording for each month and risk class the number of times the null hypothesis has been rejected, separated in over- $(+)$ and under- $(-)$ representation. The last two columns contain the number classes respectively tested, and in total ($nC$).

\begin{table}[hb]
\centering
\caption{Summary for test results: modularity}\label{tab:test_mod_sum}
\begin{tabular}{l|cc|cc|cc|ccc}
&\multicolumn{2}{c|}{L}&\multicolumn{2}{c|}{M}&\multicolumn{2}{c|}{H}\\
&  + &  - &  + &  - &  + &  - &tested& $nC_{95\%}$ &$nC$\\
\hline
Jan	&	4	&	9	&	8	&	4	&	6	&	5	&	17	&13&	1971	\\
Feb	&	5	&	11	&	9	&	2	&	9	&	7	&	20	&15&	1900	\\
Mar	&	5	&	13	&	7	&	2	&	8	&	4	&	20	&14&	2070	\\
Apr	&	4	&	9	&	7	&	3	&	7	&	6	&	19	&14&	1902	\\
May	&	3	&	9	&	8	&	2	&	7	&	5	&	18	&13&	1856	\\
Jun	&	5	&	12	&	11	&	3	&	6	&	6	&	21	&15&	2148	\\
Jul	&	6	&	8	&	10	&	5	&	5	&	3	&	17	&12&	1862	\\
Aug	&	5	&	12	&	8	&	3	&	8	&	5	&	21	&16&	2608	\\
Sep	&	3	&	9	&	9	&	2	&	4	&	4	&	16	&12&	1879	\\
Oct	&	5	&	11	&	9	&	2	&	6	&	4	&	18	&13&	1922	\\
Nov	&	5	&	9	&	7	&	4	&	5	&	4	&	17	&11&	2083	\\
Dec	&	3	&	11	&	10	&	3	&	7	&	3	&	19	&15&	2323	\\
\end{tabular}

\end{table}

\begin{table}[hp]
\centering
\caption{Summary for test results: hierarchy}\label{tab:test_ag_sum}
\begin{tabular}{l|cc|cc|cc|ccc|c}
&\multicolumn{2}{c|}{L}&\multicolumn{2}{c|}{M}&\multicolumn{2}{c|}{H}\\
&  + &  - &  + &  - &  + &  - &  tested&  $nC_{95\%}$&  $nC$ & $h$\\
\hline								
Jan	&	5	&	5	&	5	&	4	&	3	&	5	&	12	& 11&	18&0.75\\
Feb	&	5	&	5	&	4	&	4	&	4	&	5	&	12	&11&	17&0.74	\\
Mar	&	4	&	4	&	4	&	4	&	4	&	4	&	12	&11&	18&0.74	\\
Apr	&	4	&	4	&	3	&	4	&	3	&	6	&	12	&10&	15&	0.75\\
May	&	6	&	4	&	4	&	4	&	3	&	6	&	12	&13&	18&0.74	\\
Jun	&	5	&	3	&	3	&	4	&	4	&	6	&	12	&11&	17&0.75	\\
Jul	&	4	&	3	&	3	&	4	&	3	&	5	&	12	&11&	16&	0.74\\
Aug	&	6	&	4	&	3	&	5	&	3	&	5	&	14	&11&	20&0.78	\\
Sep	&	5	&	4	&	3	&	4	&	4	&	6	&	12	&10&	17&0.74	\\
Oct	&	5	&	4	&	3	&	4	&	3	&	5	&	12	&11&	17&0.73	\\
Nov	&	5	&	3	&	4	&	4	&	3	&	5	&	12	&10&	18&0.75	\\
Dec	&	4	&	3	&	3	&	4	&	3	&	7	&	12	&12&	19&0.75	\\							

\end{tabular}

\end{table}
\clearpage
\section{Classification}\label{a3:class}
\subsection{Data pre-processing}\label{a:preparation}
It is well established \cite{friedman2001elements} that rescaling/ transforming data in order them to be $\in[0,1]$ or $\in [-1,1]$ or standardised, generally improves the performance of classification, especially when different predictors have very different scale. So, before training the models we perform data preprocessing, in particular:
\begin{enumerate}[noitemsep,label=\roman*.]
\item for in- and out- degree we use quantile transformation of the logarithm of the degree. This choice is explained by the aforementioned power-law tail distribution of these quantities, and aim to avoid too scattered data; 
\item the predictors for assortativity are already $\in [0,1]$ so they do not need preprocessing;
\item the distribution of nodes into hierarchy classes is standardised, i.e each rank is shifted and rescaled to have mean 0 ad variance 1;
\item the module is the only categorical variable. The usual binary transformation would result into a new binary variable for each possible value. As we discussed before, the number of modules is very high but a small fraction of them contains almost all the nodes, so we only keep those that have more than 500 nodes and merge all the remaining into a residual class;
\item quantile transformation is applied also to the log-distribution of the size.
\end{enumerate}

\subsection{Models training and hyper-parameter optimisation}\label{a:hpo}
Models training is performed using already implemented packages: for multinomial logistic and classification trees \emph{Scikit-learn} Python package \cite{scikit-learn} has been employed, while for neural networks \emph{Keras} Python package \cite{chollet2015keras} and \emph{Tensorflow} \cite{tensorflow2015-whitepaper} have been used.
 However, during optimisation, the parameters that define the \emph{architecture} of the model, the so called hyper-parameters, remain fixed. For this reason, a common practice is to train many models using different values for these hyper-parameters and compare performance according to the chosen metric(s). A thorough discussion on this topic is beyond the scope of this paper, we refer to \cite{bergstra2012random} and related literature for detailed information.

Here we apply a simple grid search for the hyper-parameters of interest. This has been done for both 1-step and 2-steps classifiers. The metrics we employ take into account the domain specific interpretation of the risk classes. In particular we want to penalise more misclassification towards lower risk classes, i.e $M\to \bar L,\,H\to \bar M,\, H\to \bar L$\footnote{$X$ indicates the real class, while $\bar{X}$ indicates the predicted class.}, and towards \emph{distant} classes, i.e  $L\to \bar H,\, H\to \bar L$.
For this reason, beside the standard accuracy and recall, we also consider weighted scores for accuracy $ws_{\text{acc}}$, recall $ws_{\text{rec}}$, precision $ws_{\text{pr}}$, which are function of the confusion matrix $C$.
With the notation
\[C_{x,y}=\abs{\{x\to \bar{y}\}}\,,\quad C_{\cdot,y}=\sum_{x}C_{x,y}, \qquad\forall x,y\in\{L,\,M,\,H\}
\]

\[
ws_{\text{acc}}=\frac{1}{C_{\cdot,\cdot}}\sum_{x,\,y\in\{L,\,M,\,H\}}C_{x,y}P^{\text{acc}}_{x,y}\,,\qquad P^{\text{acc}}=\begin{bmatrix}
    1   & -0.25 & -0.5  \\
    -0.75 & 1 & -0.25 \\
    -1       & -0.75 & 1 
\end{bmatrix}
\]

\[
ws_{\text{rec}}=\sum_{x,\,y\in\{L,\,M,\,H\}}\frac{C_{x,y}}{C_{x,\cdot}}P^{\text{rec}}_{x,y}\,,\qquad P^{\text{rec}}=\begin{bmatrix}
    1   & -0.25 & -0.75  \\
    -0.75 & 1 & -0.25 \\
    -1       & -0.75 & 1.75 
\end{bmatrix}
\]

\[
ws_{\text{pr}}=\sum_{x,\,y\in\{L,\,M,\,H\}}\frac{C_{x,y}}{C_{\cdot,y}}P^{\text{pr}}_{x,y}\,,\qquad P^{\text{pr}}=\begin{bmatrix}
    1   & -0.25 & -0.75  \\
    -0.75 & 1 & -0.25 \\
    -1       & -0.75 & 1.75 
\end{bmatrix}
\]

For classification trees, the hyper-parameter of interest is the depth, i.e the maximum number of condition to be satisfied for classification (or the length of the longest path from root to leaves). A higher value for depth results in lower training error but may lead to over-fitting. We considered value of depth from 3 to 10. For the 1-step model, the tree with depth 6 resulted the best choice, while for the 2-steps, the best results have been attained with a depth of 9 for the first step tree and 5 for the second.  
For neural networks, the hyper-parameters of interest are the number and size of hidden layers. As before, increasing too much these values may lead to over-fitting.
In order to avoid extremely high number of parameters when adding layers, we consistently reduce their size as their number increases (intuitively, the number of parameter grows as $\prod_i \abs{l_i}$, where $\abs{l_i}$ is the size of the $i$th layer). For example, in the case of 1 (hidden) layer the number of nodes is between 10 and 100, while for two layers, it goes from 5 each to 10 each. For the 1-step model the best results are obtained with 1 layer of 50 nodes, while for the 2-steps the best choice is 2 layers of 5 nodes each for the first step and 1 layer of 10 nodes for the second.


\end{document}